\documentclass[prx,aps,twocolumn,superscriptaddress,floatfix,10pt]{revtex4-2}

\usepackage{amsmath,amssymb,amsthm}
\usepackage{mathtools}
\usepackage{graphicx}
\usepackage{booktabs}
\usepackage{array}
\usepackage{multirow}
\usepackage[table]{xcolor}
\usepackage{stmaryrd}
\usepackage{bm, tikz-cd}
\usepackage{physics}
\usepackage[colorlinks=true,hyperindex,breaklinks,
            linkcolor=blue,urlcolor=blue,citecolor=blue]{hyperref}

\newcommand{\F}{\mathbb F}
\newcommand{\Z}{\mathbb Z}

\newcommand{\CCZ}{\mathrm{CCZ}}

\newcommand{\secref}[1]{Sec.~\ref{#1}}
\newcommand{\appref}[1]{App.~\ref{#1}}
\newcommand{\eqnref}[1]{Eq.~\eqref{#1}}
\newcommand{\figref}[1]{Fig.~\ref{#1}}
\newcommand{\tabref}[1]{Tab.~\ref{#1}}

\begin{document}

\title{Logical CCZ state cultivation}

\author{Yuta Hirasaki}
\email{yutah2@illinois.edu}
\affiliation{Department of Physics, Illinois Quantum Information Science and Technology Center, and Institute of Condensed Matter Theory, The Grainger College of Engineering, University of Illinois at Urbana-Champaign, Urbana, Illinois 61801, USA}

\author{Jong Yeon Lee}
\email{jongyeon@illinois.edu}
\affiliation{Department of Physics, Illinois Quantum Information Science and Technology Center, and Institute of Condensed Matter Theory, The Grainger College of Engineering, University of Illinois at Urbana-Champaign, Urbana, Illinois 61801, USA}
\affiliation{Korea Institute for Advanced Study, Seoul 02455, South Korea}

\date{\today}

\begin{abstract}
Logical CCZ states enable Toffoli gates, which are central to arithmetic in fault-tolerant quantum computation. We introduce an end-to-end protocol for direct CCZ-state preparation using bivariate tricycle (BT) and bivariate bicycle (BB) codes. Our factory combines the cup-product CCZ gates native to BT codes with two novel ingredients, both enabled by engineering the logical coupling between BT and BB codes through projection-preserving lifts. Index-2 lifted BB ancillas enable selective $Z$-basis initialization so that the cup-product CCZ gates prepare two disjoint logical CCZ states. We transfer these states to index-3 lifted BB memories with high code distance and use decoder-based postselection to obtain high-fidelity outputs with low overhead.

Exact simulations under circuit-level noise show that our factory can reduce overhead compared with $T$-state cultivation followed by CCZ conversion. At a physical error rate of $p=0.1\%$, our smallest factory uses 252 qubits to produce two CCZ states with a joint infidelity of $2.77\times10^{-5}$. This reduces the expected spacetime volume by roughly a factor of four and the peak qubit count per CCZ by more than a factor of ten. At $p=0.01\%$, our largest factory produces two CCZ states with an estimated joint infidelity of $\sim10^{-13}$ and an acceptance rate greater than $50\%$. These results establish a resource-efficient approach to logical CCZ-state preparation across a broad range of physical error rates, from early fault-tolerant quantum computing to the high-fidelity regime needed for cryptographically relevant quantum computation.
\end{abstract}

\maketitle
\section{Introduction}

Fault-tolerant quantum computation requires protected non-Clifford operations, yet the Eastin--Knill theorem rules out a universal transversal gate set within a fixed quantum error-correcting (QEC) code~\cite{eastin2009restrictionstransversalencodedquantum}. A standard route to universality instead implements non-Clifford gates through magic-state injection, shifting the challenge to preparing high-fidelity resource states. Magic-state distillation~\cite{bravyi2005universal} converts multiple noisy resource states into fewer states of higher fidelity using Clifford operations and Pauli measurements. Suppressing faults in these operations generally requires encoding the participating logical qubits in large-distance QEC codes, resulting in substantial spacetime overhead. Despite advances in distillation protocols~\cite{PhysRevA.86.052329,PhysRevA.87.042305,Haah2017magicstate,PhysRevLett.120.050504,PhysRevLett.123.070507,PhysRevLett.111.090505,Haah2018codesprotocols,2481602.2481604,golowich2024asymptoticallygoodquantumcodes,10.1145/3717823.3718186,10.1145/3717823.3718318,xu2026distillingmagicstatesbicycle,Wills2025}, their fault-tolerant implementations, and the use of high-rate QEC codes~\cite{Gidney2019efficientmagicstate,Litinski2019magicstate,cervia2025magicstatedistillationusing,PhysRevA.95.022316,PhysRevA.95.032338}, this overhead remains a major bottleneck.

Magic-state cultivation~\cite{gidney2024magic} and zero-level distillation~\cite{thxx-njr6} pursue a complementary strategy: preparing and verifying resource states in small error-detecting codes before transferring them into higher-distance memories. By rejecting detected faults rather than correcting them during the early stages, these approaches reduce the encoding overhead at the cost of postselection. Their efficiency therefore hinges on suppressing undetected logical errors while maintaining a sufficiently high acceptance probability. Existing magic-state cultivation protocols have focused on $T$ states using small surface or colour codes~\cite{gidney2024magic,chen2025efficientmagicstatecultivation,claes2025cultivatingtstatessurface,sahay2025fold,vaknin2026efficientmagicstatecultivation,hirano2026efficientmagicstatecultivation,f7fz-6zsy}. However, for quantum arithmetic, CCZ states provide a direct resource for Toffoli gates used in addition, modular arithmetic, and oracle circuits~\cite{gidney2018halvingcostquantumaddition}. The cost of supplying these states is particularly consequential for factorization algorithms~\cite{gidney2025factor2048bitrsa, Gidney2021howtofactorbit}. The conventional route converts four cultivated $T$ states into one CCZ state~\cite{Gidney2019efficientmagicstate, PhysRevA.87.022328, PhysRevLett.118.090501}, incurring additional overhead from the conversion.

\begin{figure*}[!t]
    \centering\includegraphics[width=1.0\linewidth]{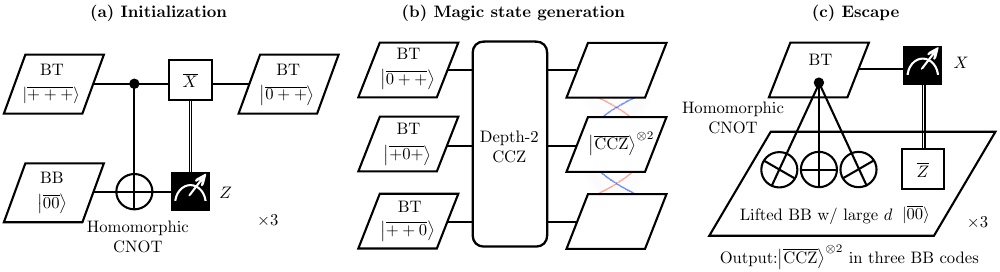}
    \caption{\textbf{Protocol overview.} 
    {\bf (a)} Three BT blocks are selectively initialized in $\ket{\overline{0++}}$, $\ket{\overline{+0+}}$, and $\ket{\overline{++0}}$ using homomorphic measurements with index-2 lifted BB codes. 
    {\bf (b)} A depth-two cup-product circuit acting on the properly initialized BT states directly generates two disjoint logical CCZ states.  
    {\bf (c)} Homomorphic CNOTs subsequently teleport these states into three higher-distance BB blocks constructed through index-3 lifts.
    }
    \label{Fig:CultivationOverview}
\end{figure*}

Here we develop an end-to-end CCZ-state factory based on bivariate tricycle and bicycle (BT/BB) codes. BT codes support constant-depth cup-product logical CCZ gates~\cite{breuckmann2024cups,ghhp-cytl, jacob2026singleshotdecodingfaulttolerantgates}, and Ref.~\cite{li2026transversaldimensionjumpproduct} identified small BT/BB code pairs for native magic-state generation. However, a cup-product CCZ gate alone does not constitute a usable resource-state factory. Applying the gate to logical $\ket{\overline{+}}$ states produces a hypergraph state~\cite{Rossi_2013} of nine logical qubits rather than independent CCZ states~\cite{ghhp-cytl,li2026transversaldimensionjumpproduct}. Moreover, the prepared states remain in a low-distance code, requiring a protected \emph{escape} into higher-distance memory before they can be reliably stored or used.

We resolve both obstacles by engineering the logical coupling through projection-preserving lifts of BB codes~\cite{hirasaki2026liftingliftedproductcodes}. First, an index-2 lift yields a rank-one logical image, enabling \emph{selective logical initialization} through homomorphic measurements~\cite{huang2023homomorphic,PhysRevX.15.021065}. Initializing one logical qubit in each of three BT blocks in the $\ket{0}$ state allows the native depth-two cup-product circuit to directly generate two disjoint logical CCZ states. Second, an index-3 lift preserves the full rank-two logical image of the remaining degrees of freedom while increasing the code distance, enabling their \emph{escape} into higher-distance BB memories through homomorphic CNOTs.

We use the term \emph{cultivation} in a broad sense: high-fidelity CCZ states are prepared in small codes operating in error-detection mode and subsequently transferred to larger-distance memories with low spacetime overhead. Unlike the conventional approach~\cite{gidney2024magic}, our protocol generates the resource directly through non-Clifford logical gates, at \emph{zero-level}~\cite{thxx-njr6, itogawa2026zerolevelcczdistillation}, without successive rounds of logical-state verification~\cite{Chamberland2020}.

We perform exact simulations~\cite{fby6-xjbm, kim2026xtim} of our factory under a circuit-level noise model motivated by neutral-atom hardware. We evaluate the acceptance rate, output infidelity, and expected spacetime volume, and benchmark our protocol against a state-of-the-art approach based on $T$-state cultivation~\cite{sahay2025fold}. Our results demonstrate that direct CCZ-state preparation achieves a severalfold reduction in spacetime overhead compared with the $T$-state-based approach.

\vspace{5pt}
\noindent
\textbf{Related work.}
Cup-product logical CCZ gates on qLDPC codes have been developed in Refs.~\cite{breuckmann2024cups,hsin2025classifyinglogicalgatesquantum,breuckmann2026cupsgatesiioperads,wcxs-w69t,zhu2025topologicaltheoryqldpcnonclifford,zhu2025transversalnoncliffordgatesqldpc,lin2024transversalnoncliffordgatesquantum,li2026theorycohomologicalinvariantsquantum,tiew2026copycupgatestensorproducts,jacob2026singleshotdecodingfaulttolerantgates,ghhp-cytl,li2026transversaldimensionjumpproduct,tan2025singleshotuniversalityquantumldpc}. Native non-Clifford operations and postselection have also been explored for magic-state preparation~\cite{ghhp-cytl,PhysRevResearch.7.023080,li2026transversaldimensionjumpproduct,dck4-x9c2,f7fz-6zsy,min2026codedcliffordmeasurementsmultiqubit}. Our contribution is to integrate selective logical initialization, native CCZ generation, and high-distance escape into an explicit cultivation protocol, with circuit-level simulations and resource accounting for the complete factory.
The protocols in Refs.~\cite{itogawa2026zerolevelcczdistillation, min2026codedcliffordmeasurementsmultiqubit, min2026directcultivationentangledcsrangle} have lower fault distances and lower output fidelities.

\section{Protocol and results}

\begin{table*}[t!]
\centering
\begingroup
\fontsize{8}{12}\selectfont
\renewcommand{\arraystretch}{1.35}
\setlength{\extrarowheight}{1pt}
\begin{ruledtabular}
\setlength{\tabcolsep}{2pt}
\begin{tabular}{ccccccccccc}
\shortstack{Factory\\$(n_{\mathrm{3D}},d_f)$} & \shortstack{Space\\overhead} & \shortstack{Gate\\depth} & $p$ & \shortstack{Online\\AR} & \shortstack{Joint infidelity\\$1-F_{\mathrm{joint}}$} & \shortstack{CCZ1 infidelity\\$1-F_1$} & \shortstack{CCZ2 infidelity\\$1-F_2$} & \shortstack{Failure rate\\${N_{\mathrm{fail}}}/{N_{\mathrm{accept}}}$} & $\overline V_{\CCZ}$ & $\overline V_{4T\to\CCZ}$\\
\hline
\multirow{2}{*}{$(21,3)$} & \multirow{2}{*}{252} & \multirow{2}{*}{80} & $0.1\%$ & $44.05\%$ & $2.77\times10^{-5}$ & $2.02\times10^{-5}$ & $1.95\times10^{-5}$ & $96/3{,}161{,}161$ & $1.48\times10^{4}$ & $5.80\times10^{4}$\\
 & & & $0.01\%$ & $91.06\%$ & $6.04\times10^{-8}$ & $4.25\times10^{-8}$ & $3.94\times10^{-8}$ & $62/881{,}951{,}834$ & $9.12\times10^{3}$ & --\\
\hline\hline
\multirow{2}{*}{$(27,3)$} & \multirow{2}{*}{324} & \multirow{2}{*}{75} & $0.1\%$ & $35.71\%$ & $1.83\times10^{-5}$ & $1.54\times10^{-5}$ & $1.23\times10^{-5}$ & $91/4{,}693{,}422$ & $2.18\times10^{4}$ & $5.88\times10^{4}$\\
 & & & $0.01\%$ & $89.04\%$ & $3.25\times\!10^{-8}$ & $2.87\times10^{-8}$ & $2.40\times10^{-8}$ & $30/854{,}814{,}901$ & $1.16\times10^{4}$ & --\\
\hline\hline
\multirow{2}{*}{$(36,4)$} & \multirow{2}{*}{432} & \multirow{2}{*}{84} & $0.1\%$ & $16.95\%$ & $2.61\times10^{-7}$ & $1.06\times10^{-7}$ & $2.04\times10^{-7}$ & $24/91{,}763{,}021$ & $4.39\times10^{4}$ & $1.27\times10^{5}$\\
 & & & $0.01\%$ & $82.78\%$ & -- & -- & -- & -- & $1.68\times10^{4}$ & --\\
\hline\hline
\multirow{2}{*}{$(81,5)$} & \multirow{2}{*}{972} & \multirow{2}{*}{83} & $0.1\%$ & $3.11\%$ & $<5.5\times10^{-8}$ & -- & -- & $\bm{0}/55{,}081{,}056$ & $8.62\times10^{5}$ & --\\
 & & & $0.01\%$ & $69.36\%$ & -- & -- & -- & -- & $4.42\times10^{4}$ & --\\
\end{tabular}
\end{ruledtabular}
\endgroup
\caption{\label{tab:factory-benchmark}\textbf{Performance of the CCZ-state cultivation factory.}
$n_{\mathrm{3D}}$ is the number of data qubits in one BT block and $d_f$ is the target factory fault distance.
The error rates of single-qubit gates, single-qubit resets/measurements and two-qubit CZ gates, and CCZ gates are $p/10$, $p$, and $5p$, respectively. Space overhead is the peak physical-qubit count, including ancillas. The online acceptance rate (AR) is conditional on successful preparation of all three BT blocks and their initialization BB blocks before the homomorphic CNOTs. $1-F_{\mathrm{joint}}$, $1-F_1$, and $1-F_2$ are the mean joint and individual infidelities over accepted attempts; $N_{\mathrm{fail}}/N_{\mathrm{accept}}$ gives the fraction of corrupted final states within accepted attempts. The expected spacetime volume $\overline V_{\CCZ}$ per CCZ includes ancillary-code preparation. The definition of expected spacetime volume $\overline V_{\CCZ}$ per CCZ follows Ref.~\cite{sahay2025fold}, and is described in \appref{App:ResourceEstimate}. The last column gives the $T$-cultivation-plus-conversion volume at $p=0.1\%$ with a comparable output CCZ fidelity; see \appref{App:ResourceEstimate} for more details.
}
\end{table*}

Our cultivation protocol, illustrated in \figref{Fig:CultivationOverview}, extends the hypergraph-state preparation scheme of Ref.~\cite{li2026transversaldimensionjumpproduct} into an end-to-end CCZ-state cultivation factory with two new ingredients: \emph{selective initialization} [Fig.~\ref{Fig:CultivationOverview}(a)] and  \emph{lifting-based escape} [Fig.~\ref{Fig:CultivationOverview}(c)]. We develop a postselection strategy that enables the preparation of high-fidelity CCZ states at low overhead, and benchmark the performance under circuit-level noise. Below, we describe the three stages of the protocol and evaluate its performance numerically, with circuit-level implementation and postselection detailed in \appref{App:cultivationdetails}.

\subsection{Protocol}
Our BT codes support a cup-product gate~\cite{li2026transversaldimensionjumpproduct,jacob2026singleshotdecodingfaulttolerantgates}, which we use to prepare CCZ states. Its physical implementation is specified by a trilinear form $f_{\cup}: C^1\times C^1\times C^1\to \F_2$:
% in \eqnref{eq:cup_trilinear},
\begin{align} \label{eq:CCZ_cup}
U_{\cup}
=\prod_{q,r,s}
\CCZ(q,r,s)^{f_{\cup}(e_q,e_r,e_s)},
\end{align}
where $e_q$ is the basis cochain associated with physical qubit $q$, and $q,r,s$ label qubits in three distinct BT blocks. The form $f_{\cup}$ descends to cohomology~\cite{breuckmann2024cups}, inducing the logical operation
\begin{align}
\overline{U}_{\cup}
=\prod_{i,j,k}
\overline{\CCZ}(i,j,k)^{T_{ijk}},
\label{Eq:logicalaction}
\end{align}
where $i, j, k$ label logical qubits in each code block, and $T_{ijk}\in\{0,1\}$ specifies whether the corresponding logical CCZ gate is present. For these BT codes, the cyclic and anticyclic permutations of the three physical qubit sectors give two layers of disjoint CCZ gates~\cite{li2026transversaldimensionjumpproduct}. Every physical qubit participates in exactly one gate per layer, so $U_{\cup}$ has depth two in physical CCZ gates. In a canonical logical basis $\{\alpha,\beta,\gamma\}$, $T_{ijk}$ is nonzero precisely when $i,j,k$ are pairwise distinct, giving six logical CCZ interactions across the three blocks~\cite{jacob2026singleshotdecodingfaulttolerantgates}.

\vspace{5pt} 
\noindent
\textbf{(a) Initialization.}
The key to generating two disjoint CCZ states is to selectively initialize one logical qubit in each BT block prepared in the $\ket{\overline{+++}}$ state, so that $\overline{U}_\cup$ in Eq.~\eqref{Eq:logicalaction} generates two distinct CCZ states. If we prepare $\alpha_1$, $\beta_2$, and $\gamma_3$ in the logical $\ket{\overline0}$ state, four of the six logical CCZ actions become trivial, leaving precisely two disjoint logical CCZ gates acting on $(\beta_1, \gamma_2, \alpha_3)$ and $(\gamma_1, \alpha_2, \beta_3)$. 

Graph surgery~\cite{Williamson2026} provides a general route to this selective initialization, but $Z$-surgery on the low $Z$-distance BT codes introduces either substantial logical errors in error correcting mode or postselection overhead in error detecting mode. Instead, we use index-2 lifted BB ancillas with a rank-one logical map to selectively measure $\overline{Z}_{\alpha_1}$, $\overline{Z}_{\beta_2}$, and $\overline{Z}_{\gamma_3}$ through homomorphic measurements~\cite{huang2023homomorphic,PhysRevX.15.021065}. Each ancilla has a rank-one logical map whose image contains only the targeted BT logical-$Z$ operator. 

We first prepare each BT block in $\ket{\overline{+++}}$ by initializing its physical qubits in $\ket{+}$ and measuring the $Z$ checks once. The meta-$Z$ checks provide redundant syndrome constraints for detecting measurement errors and rejecting unreliable preparations. Each BB ancilla is prepared from physical $\ket{0}$ states using $X$-check measurements, coupled to its BT block through a homomorphic CNOT, and transversally measured in the $Z$ basis. The decoded readout yields the targeted logical-$Z$ outcome, and conditional logical-$X$ feedback prepares the three codes in
\begin{align}
\ket{\overline{0++}}_1
\otimes\ket{\overline{+0+}}_2
\otimes\ket{\overline{++0}}_3.
\label{eq:selectiveinput}
\end{align}

\vspace{5pt}
\noindent
\textbf{(b) Magic state generation.}
We apply the cup-product logical CCZ gate to the three BT code blocks. With the initialization described above, the native depth-two physical circuit implements exactly the two logical CCZ gates
\begin{align}\label{eq:twoccz}
 \overline{\CCZ}(\beta_1,\gamma_2,\alpha_3)\,\overline{\CCZ}(\gamma_1,\alpha_2,\beta_3),
\end{align}
producing two disjoint logical CCZ states hosted in three BT code blocks. The physical depth is 2 for all the BT codes studied here.

\vspace{5pt} 
\noindent
\textbf{(c) Escape.}
The prepared state is transferred to index-3 lifted BB codes with sufficient distance to store it. We apply a homomorphic CNOT from the BT code to the BB code, measure the BT code transversally in the $X$ basis, and apply logical-frame feedback to the output BB code. The escape CNOTs can spread a single fault into correlated errors, which are carefully handled in postselection. The correlated CZ errors triggered by CCZ gates are also handled in postselection. We perform three rounds of syndrome extraction (SE) and decoding after the escape. Details of decoding and postselection are discussed in \appref{App:cultivationdetails}.

\subsection{Benchmark}
We simulate the factory exactly under a circuit-level noise model motivated by neutral-atom experiments~\cite{sahay2025fold}. Depolarizing noise follows single-qubit, CZ and CCZ gates with probabilities $p/10$, $p$ and $5p$, respectively. Resets and measurements have bit-flip errors with probability $p$, and idling is noiseless. Each CNOT is implemented with a CZ gate and two Hadamard gates.

The results in \tabref{tab:factory-benchmark} are grouped by block size $n_{\mathrm{3D}}$ and target fault distance $d_f$ at $p=0.1\%$ and $0.01\%$. The joint fidelity $F_{\mathrm{joint}}$ is the overlap of the two-output state with $\ket{\overline\CCZ}^{\otimes 2}$. The individual fidelities $F_1$ and $F_2$ are the overlaps of the reduced states with the ideal CCZ state. We also report the fraction of accepted pairs with nonunit joint fidelity. For the $d_f=5$ factory at $p=0.1\%$, no failure is observed among the $55{,}081{,}056$ accepted pairs, giving a one-sided 95\% upper bound of $5.5\times10^{-8}$ on the joint infidelity. At $p=0.01\%$, joint infidelity for $n_{\mathrm{3D}}=36$ and $81$ is too small to sample failures so we leave them blank. For the two $d_f = 3$ factories, lowering $p$ to $0.01\%$ reduces joint infidelity by around three orders of magnitude. Assuming $p^{d_f}$ scaling, we estimate the output infidelity for $d_f = 4$ and $d_f = 5$ as $\sim 10^{-11}$ and $\sim 10^{-13}$, respectively.

At $p=0.1\%$, we compare the expected volume per accepted CCZ with that of surface-code $T$-state cultivation~\cite{sahay2025fold} followed by $4T$-to-CCZ conversion. We divide the joint infidelity by the two outputs per pair and match $(1-F_{\mathrm{joint}})/2$ to the first-order $T$-route infidelity $4(1-F_T)$, where $F_T$ is the fidelity of a cultivated $T$ state. This accounting reflects the number of CCZ states that can be supplied per logical failure event. We note that correlated output errors can cause the individual infidelities to differ from this normalized joint infidelity as shown in the table and this must be accounted for in magic state distillation~\cite{gong2026magicstatedistillationcodes}.

Our factories use less expected spacetime volume than the $T$-state based factory. For instance, the $n_{\mathrm{3D}}=21$ factory uses $1.48\times10^4$ qubit-steps per CCZ, about a quarter of the $T$-route cost of $5.80\times10^4$. The conversion alone requires $1{,}348$ or $1{,}796$ physical qubits at surface-code distances 13 and 15, respectively. All four direct factories have smaller peak qubit counts. Lowering $p$ to $0.01\%$ increases the acceptance rate and further reduces the expected volumes for our factories. We describe the resource accounting in \appref{App:ResourceEstimate}.

\section{Codes and Logical Coupling} \label{sec:code}

\begin{figure*}[t!]
    \centering
    \includegraphics[width=1.0\linewidth]{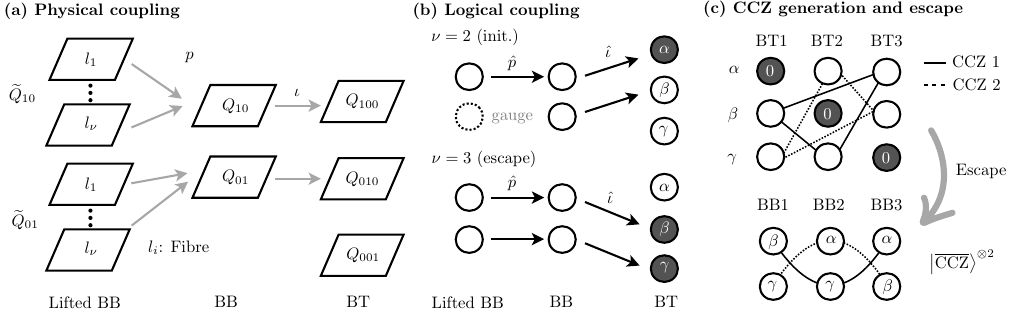}
    \caption{\textbf{Lifting-based mechanisms.} The coupling among the BT code $Q_{\bullet}^\mathrm{ABC}$, the BB code $Q_{\bullet}^\mathrm{AB}$, and the lifted BB code $\widetilde{Q}_{\bullet}^\mathrm{AB}$ is illustrated.
    {\bf (a)} 
    Physical coupling of the three codes. We use the shorthand notation $\widetilde{Q}_{ij}=\widetilde{A}_i\otimes \widetilde{B}_j$, $Q_{ij}=A_i\otimes B_j$, and $Q_{ijk}=A_i\otimes B_j\otimes C_k$.
    The projection $p$ maps the $\nu$ sheets of the lifted BB code $\widetilde{Q}_\bullet^{AB}$ onto the base BB code $Q_\bullet^{AB}$.
    The inclusion $\iota$ embeds its two qubit sectors into the $Q_{100}$ and $Q_{010}$ sectors of the BT code $Q_\bullet^{ABC}$.
    Their composition $f=\iota\circ p$ specifies the homomorphic CNOT. 
    {\bf (b)} Induced logical couplings. The circles denote logical qubits, with the BT logical qubits labelled by $\alpha$, $\beta$, and $\gamma$.
    The chosen index-2 $AB$ lift has logical image $\langle\alpha\rangle$, enabling selective logical-$Z$ initialization through homomorphic measurement; its uncoupled logical qubit is treated as a gauge degree of freedom. The index-3 $BC$ lift preserves both base BB logical degrees of freedom and couples to $\langle\beta,\gamma\rangle$ for escape. Shaded circles identify the coupled BT logical qubits.
    {\bf (c)} The logical coupling is engineered to initialize $\alpha_1,\beta_2,\gamma_3$ in $\ket{\overline{0}}$, with the remaining logical qubits in $\ket{\overline{+}}$. This reduces the native cup-product action to two disjoint CCZ-state preparations, indicated by solid and dashed connections. The two logical qubits hosting the CCZ states in each block are transferred to the escape BB code.}
    \label{Fig:LogicalCoupling}
\end{figure*}

We describe the BB/BT codes used in our factory and their physical and logical couplings. Figure~\ref{Fig:LogicalCoupling} illustrates the couplings between the codes, and \tabref{tab:codes} summarizes their parameters. In \secref{Sec:BBBTreview}, we first review the BB/BT codes~\cite{li2026transversaldimensionjumpproduct}, along with chain maps and homomorphic CNOTs. We then discuss the BB codes constructed through lifting~\cite{hirasaki2026liftingliftedproductcodes} and their couplings to the BT codes in \secref{Subsec:liftBB}. We also perform memory experiments on the escape codes to verify that they provide sufficient logical error suppression to store the output magic states.

\subsection{Review of BB/BT codes}
\label{Sec:BBBTreview}
BB and BT codes are tensor-product chain complexes over the group algebra $R=\F_2[\Z_l\times \Z_m]$. Let $A_\bullet, B_\bullet$ and $C_\bullet$ be length-one chain complexes:
\begin{align}\label{Eq:BaseClassicalChain}
    A_1\xrightarrow{a} A_0, \quad B_1\xrightarrow{b}B_0,\quad C_1\xrightarrow{c}C_0,
\end{align}
where each module is a copy of $R$ and the boundary maps are given by multiplication by elements $a,b,c \in R$, respectively.
The tensor product of the first two chain complexes defines the BB code:
\begin{equation}
    \begin{tikzcd}[column sep=1em] 
    Q^{\mathrm{AB}}_0
      & 
      & A_0 \otimes B_0
      & \\
    Q^{\mathrm{AB}}_1 \arrow[u, "H^{\mathrm{AB}}_X"]
      & A_1 \otimes B_0 \arrow[ru, "a \otimes 1"]
      &
      & A_0 \otimes B_1 \arrow[lu, "1 \otimes b"'] \\
    Q^{\mathrm{AB}}_2 \arrow[u, "(H^{\mathrm{AB}}_Z)^T"]
      &
      & A_1 \otimes B_1
          \arrow[lu, "1 \otimes b"]
          \arrow[ru, "a \otimes 1"']
      &
    \end{tikzcd}\label{Eq:2Dchain}.
\end{equation}
$Q_2^\mathrm{AB} = A_1\otimes B_1$ corresponds to the $Z$-checks, $Q_1^\mathrm{AB} = A_1\otimes B_0\oplus A_0\otimes B_1$ to physical qubits, and $Q_0^\mathrm{AB} = A_0\otimes B_0$ to $X$-checks. Logical $Z$ operators are cycles modulo boundaries, $H_1=\ker\partial_1/\operatorname{im}\partial_2$, and logical $X$ operators belong to the dual cohomology $H^1$.

The tensor product of all three chain complexes gives the BT code:
\begin{equation}
\begin{tikzcd}[column sep=0.2em]
Q^{\mathrm{ABC}}_0 &  & {\scriptstyle A_0\otimes B_0\otimes C_0} & \\
Q^{\mathrm{ABC}}_1 \arrow[u, "H^{\mathrm{ABC}}_X"] & {\scriptstyle A_1\otimes B_0\otimes C_0}  \arrow[ru]  & {\scriptstyle A_0\otimes B_1\otimes C_0}  \arrow[u]  & {\scriptstyle A_0\otimes B_0\otimes C_1}  \arrow[lu]  \\
Q^{\mathrm{ABC}}_2 \arrow[u, "(H^{\mathrm{ABC}}_Z)^T"]  & {\scriptstyle A_1\otimes B_1\otimes C_0}  \arrow[u] \arrow[ru] & {\scriptstyle A_1\otimes B_0\otimes C_1}  \arrow[lu] \arrow[ru] & {\scriptstyle A_0\otimes B_1\otimes C_1}  \arrow[u] \arrow[lu] \\
Q^{\mathrm{ABC}}_3 \arrow[u, "(M^{\mathrm{ABC}}_Z)^T"] & & {\scriptstyle A_1\otimes B_1\otimes C_1} \arrow[lu] \arrow[u] \arrow[ru] &                                                    
\end{tikzcd}.
\label{eq:3D_complex}
\end{equation}
$Q_3^\mathrm{ABC}$ corresponds to the meta-$Z$ checks, with $Q_2^\mathrm{ABC},Q_1^\mathrm{ABC},Q_0^\mathrm{ABC}$ again representing $Z$-checks, qubits, and $X$-checks. The length-three structure provides an additional \emph{meta-$Z$ check}: an ideal $Z$-syndrome $s_Z$ satisfies $M_Z^{\mathrm{ABC}}s_Z=0$, allowing measurement errors in the $Z$-checks to be identified.

All the BT codes studied here encode three logical qubits. Here, we take the canonical basis $\alpha, \beta, \gamma$ to discuss the logical actions of homomorphic CNOTs: Appendix~\ref{app:logical-representatives} provides the explicit logical representatives for each BT code. For brevity, we use $\langle \cdots \rangle$ to denote the logical-$Z$ space corresponding to the labels in $(\cdots)$.

\vspace{5pt}
\noindent
\textbf{Native CCZ gate.}
In BT codes, the cup product defines a trilinear form $f_\cup$ on 1-cochains satisfying
\begin{align}
f_\cup(b,u,v) =f_\cup(u,b,v) =f_\cup(u,v,b)=0
\label{eq:cup_trilinear}
\end{align}
for any $b\in\operatorname{im}H_X^T$ and $u,v\in\ker H_Z$~\cite{breuckmann2024cups, ghhp-cytl}. This property ensures that the physical CCZ circuit $U_\cup$ in \eqnref{eq:CCZ_cup} preserves the codespace. The logical action is given by Eq.~\eqref{Eq:logicalaction}, with
\begin{align}
T_{ijk}
=f_\cup(\overline X_i,\overline X_j,\overline X_k),
\label{eq:logical-cup-tensor}
\end{align}
where $\overline X_i,\overline X_j,\overline X_k$ denote the binary supports of logical $X$ representatives in the respective blocks. By Eq.~\eqref{eq:cup_trilinear}, $T_{ijk}$ is independent of the choice of representatives. For the BT codes studied here, $T_{ijk}$ vanishes unless $i,j,k$ are pairwise distinct, yielding a product of six logical CCZ gates~\cite{jacob2026singleshotdecodingfaulttolerantgates} as shown in \figref{Fig:LogicalCoupling}(c).

\vspace{5pt}
\noindent
\textbf{BB/BT pairing.} The BB and BT codes constructed above are related by an inclusion chain map $\iota$, built as follows~\cite{li2026transversaldimensionjumpproduct}. Consider the chain complex $C'_\bullet$
\begin{align}
 C'_\bullet:  C'_1 = \{0\}\xlongrightarrow{c' = 0} C'_0=R.
\end{align}
The $R$-linear maps ${\iota'_1}=0$ and ${\iota'_0}:C'_0\to C_0$, $r\mapsto r$, form a chain map ${\iota'_\bullet}:C'_\bullet\to C_\bullet$,
\begin{equation}\label{eq:componentmap}
    \begin{tikzcd}
        {C_1}  & {C_0}\\
        {C_1'} & {C_0'}
        \arrow["c", from=1-1, to=1-2]
        \arrow["{\iota'_1}", from=2-1, to=1-1]
        \arrow["{\iota'_0}", from=2-2, to=1-2]
        \arrow["c'", from=2-1, to=2-2]
    \end{tikzcd}.
\end{equation}
Tensoring $\iota'_\bullet$ with the identity on the $A$ and $B$ factors gives the inclusion $\iota:Q^{AB}_\bullet\to Q^{ABC}_\bullet$,
\begin{align}
 \iota_q(z)&=z\otimes 1,\quad q=0,1,2,\nonumber\\
 z&\in Q_q^{AB}=\bigoplus_{r+s=q}A_r\otimes_R B_s.
\end{align}
In degree one, the two BB summands map to the first two BT summands, $A_1\otimes B_0\otimes C_0$ and $A_0\otimes B_1\otimes C_0$, leaving $A_0\otimes B_0\otimes C_1$ decoupled. This gives the commuting diagram
\begin{equation}\label{Eq:Inclusion}
    \begin{tikzcd}
        {Q_3^{\mathrm{ABC}}}  & {Q_2^{\mathrm{ABC}}} & {Q_1^{\mathrm{ABC}}} & {Q_0^{\mathrm{ABC}}}\\
        {} & {Q_2^{\mathrm{AB}}} & {Q_1^{\mathrm{AB}}} & {Q_0^{\mathrm{AB}}}
        \arrow[from=1-1, to=1-2]\arrow[from=1-2, to=1-3]\arrow[from=1-3, to=1-4]
        \arrow[from=2-2, to=2-3]\arrow[from=2-3, to=2-4]
        \arrow["{\iota_2}", from=2-2, to=1-2]
        \arrow["{\iota_1}", from=2-3, to=1-3]
        \arrow["{\iota_0}", from=2-4, to=1-4]
    \end{tikzcd}.
\end{equation}
The same construction applies to $Q^{BC}_\bullet$ and $Q^{AC}_\bullet$. A chain map induces a \emph{homomorphic CNOT} circuit~\cite{huang2023homomorphic, PhysRevX.15.021065} between the two codes, as described below.

\vspace{5pt}
\noindent
\textbf{Homomorphic CNOT.} Consider two codes $Q_\bullet$ and $\widetilde{Q}_\bullet$ related by chain maps
\begin{equation}
    \begin{tikzcd}
         {Q_2} & {Q_1} & {Q_0}\\
        {\widetilde{Q}_2} & {\widetilde{Q}_1} & {\widetilde{Q}_0}
        \arrow[from=1-1, to=1-2]\arrow[from=1-2, to=1-3]
        \arrow[from=2-2, to=2-3]\arrow[from=2-1, to=2-2]
        \arrow["{f_2}", from=2-1, to=1-1]
        \arrow["{f_1}", from=2-2, to=1-2]
        \arrow["{f_0}", from=2-3, to=1-3]
    \end{tikzcd}.
\end{equation}
The homomorphic CNOT is the product of physical CNOTs with control qubits $i$ in $Q_1$ and target qubits $j$ in $\widetilde Q_1$ for which $(f_1)_{ij}=1$. Its action on physical Paulis is
\begin{align}
 X_Q(x)&\longmapsto X_Q(x)X_{\widetilde Q}(f_1^Tx),\nonumber\\
 Z_{\widetilde Q}(z)&\longmapsto Z_Q(f_1z)Z_{\widetilde Q}(z).
\end{align}
The other two Pauli types, $Z_Q$ and $X_{\widetilde Q}$, are unchanged. The chain-map identities $\partial_1f_1=f_0\widetilde\partial_1$ and $f_1\widetilde\partial_2=\partial_2f_2$ ensure that this circuit preserves the code space~\cite{huang2023homomorphic, PhysRevX.15.021065}.

Let $\hat f_1:H_1(\widetilde Q_\bullet)\to H_1(Q_\bullet)$ be the induced map on logical $Z$ classes. In paired logical $X/Z$ bases, its matrix determines the logical action:
\begin{align}
\overline X_i^Q&\longmapsto\overline X_i^Q\prod_a(\overline X_a^{\widetilde Q})^{(\hat f_1)_{ia}},\nonumber\\
\overline Z_a^{\widetilde Q}&\longmapsto\prod_i(\overline Z_i^Q)^{(\hat f_1)_{ia}}\overline Z_a^{\widetilde Q}.
\label{eq:logical-cnot-action}
\end{align}
The logical operators $\overline Z_i^Q$ and $\overline X_a^{\widetilde Q}$ are unchanged. This action enables logical measurements and logical state teleportation.

For logical measurements, initialize $\widetilde Q_\bullet$ in $\ket{\overline 0}^{\otimes\tilde k}$. The logical-$Z$ operators of this prepared state transform as
\begin{align}\label{Eq:HCNOTtransformation}
 \overline Z_a^{\widetilde Q} = 1
 &\longmapsto L_a^Q\overline Z_a^{\widetilde Q} = 1,
 &L_a^Q&=\prod_i(\overline Z_i^Q)^{(\hat f_1)_{ia}}.
\end{align}
Transversal $Z$ readout of $\widetilde Q_1$ gives its $Z$-check outcomes and logical-$Z$ outcomes $(-1)^{m_a}$. From Eq.~\eqref{Eq:HCNOTtransformation}, we obtain $L_a^Q = (-1)^{m_a}$. Thus the measured logical-$Z$ subspace is $\operatorname{im}\hat f_1$. Dually, preparing $Q$ in logical $+$ states and reading it out in $X$ measures the logical-$X$ subspace $\operatorname{im}\hat f^1$ of $\widetilde Q$, where $f^1=f_1^T$ induces $\hat f^1:H^1(Q_\bullet)\to H^1(\widetilde Q_\bullet)$.

For logical state teleportation from $Q_\bullet$ to $\widetilde Q_\bullet$, initialize $\widetilde Q_\bullet$ in logical $\ket{\overline 0}^{\otimes\tilde k}$ states, apply the homomorphic CNOT, transversally measure $Q_1$ in $X$, and apply adaptive logical-$Z$ feedback. In this way, the logical subspace $\operatorname{im}\hat{f}_1$ in $Q_\bullet$ is faithfully transferred to the logical subspace in $\operatorname{im}\hat{f}^1$.

Let us return to the inclusion map in Eq.~\eqref{Eq:Inclusion}. For the codes studied here, the induced map $\hat{\iota}_1:H_1(Q^{\mathrm{AB}}_\bullet)\to H_1(Q^{\mathrm{ABC}}_\bullet)$ is injective~\cite{li2026transversaldimensionjumpproduct}, which allows faithful logical state transfer between the BB and BT codes. Table~\ref{tab:codes} gives the images of the transfer maps. See \figref{Fig:LogicalCoupling} for the illustration of coupling.

\begin{table*}[t!]
    \centering
    \begin{ruledtabular}
    \setlength{\tabcolsep}{3pt}
        \begin{tabular}{cllccccl}
            $(n_{\mathrm{3D}},d_f)$ & Role & Code parameters & $w_X/w_Z$ & $q_X/q_Z$ & $d_m$ & SE CNOT depth & Logical image\\
            \hline
            \multirow{6}{*}{$(21,3)$}& \multirow{2}{*}{Base BB} & \multirow{2}{*}{$\llbracket14,2,(3,3)\rrbracket$} & \multirow{2}{*}{$4/4$} & \multirow{2}{*}{$2/2$} & \multirow{2}{*}{--} & \multirow{2}{*}{$(4,4,4)$} & $\langle\alpha+\beta,\alpha+\gamma\rangle$ (AB)\\
             & & & & & & & $\langle\alpha,\gamma\rangle$ (BC); $\langle\alpha,\beta\rangle$ (CA)\\
             & BT for CCZ & $\llbracket21,3,(6,3)\rrbracket$ & $6/4$ & $2/4$ & 3 & $(6,4,8)$ & --\\
            & BB init. 1 & $\llbracket28,1,1;(8,3)\rrbracket$ & $6/6$ & $\{2,4\}/\{2,4\}$ & -- & $(6,6,6)$ & $\langle\alpha\rangle$\\
            & BB init. 2/3 & $\llbracket28,1,1;(6,6)\rrbracket$ & $6/6$ & $\{2,4\}/\{2,4\}$ & -- & $(6,6,6)$ & $\langle\beta\rangle;\langle\gamma\rangle$\\
            & Escape BB & $\llbracket42,2,(8,8)\rrbracket$ & $8/8$ & $4/4$ & -- & $(8,8,8)$ & $\langle\beta,\gamma\rangle;\langle\gamma,\alpha\rangle;\langle\alpha,\beta\rangle$\\
            \hline

            \multirow{4}{*}{$(27,3)$}& Base BB & $\llbracket18,2,(3,3)\rrbracket$ & $4/4$ & $2/2$ & -- & $(4,4,4)$ & $\langle\alpha,\beta\rangle;\langle\beta,\gamma\rangle;\langle\gamma,\alpha\rangle$\\
             & BT for CCZ & $\llbracket27,3,(6,3)\rrbracket$ & $6/4$ & $2/4$ & 3 & $(6,4,8)$ & -\\
            & BB init. & $\llbracket36,1,1;(6,3)\rrbracket$ & $4/4$ & $2/2$ & -- & $(4,4,4)$ & $\langle\alpha\rangle;\langle\beta\rangle;\langle\gamma\rangle$\\
            & Escape BB & $\llbracket54,2,(9,9)\rrbracket$ & $8/8$ & $4/4$ & -- & $(8,8,8)$ & $\langle\beta,\gamma\rangle;\langle\gamma,\alpha\rangle;\langle\alpha,\beta\rangle$\\
            \hline

            \multirow{5}{*}{$(36,4)$}& \multirow{2}{*}{Base BB} & \multirow{2}{*}{$\llbracket24,2,(4,4)\rrbracket$} & \multirow{2}{*}{$4/4$} & \multirow{2}{*}{$2/2$} & \multirow{2}{*}{--} & \multirow{2}{*}{$(4,4,4)$} & $\langle\alpha+\beta,\alpha+\gamma\rangle$ (AB)\\
             & & & & & & & $\langle\beta,\alpha+\gamma\rangle$ (BC); $\langle\alpha,\gamma\rangle$ (CA)\\
             & BT for CCZ & $\llbracket36,3,(8,4)\rrbracket$ & $6/4$ & $2/4$ & 4 & $(6,4,8)$ & --\\
            & BB init. & $\llbracket48,1,1;(8,10)\rrbracket$ & $8/8$ & $4/4$ & -- & $(8,8,8)$ & $\langle\alpha\rangle;\langle\beta\rangle;\langle\gamma\rangle$\\
            & Escape BB & $\llbracket72,2,(12,12)\rrbracket$ & $8/8$ & $4/4$ & -- & $(8,8,8)$ & $\langle\beta,\gamma\rangle;\langle\gamma,\alpha\rangle;\langle\alpha,\beta\rangle$\\
            \hline

            \multirow{4}{*}{$(81,5)$}& {Base BB} & {$\llbracket54,2,(6,6)\rrbracket$} & {$4/4$} & {$2/2$} & {--} & {$(4,4,4)$} & {$\langle\alpha,\beta\rangle;\langle\beta,\gamma\rangle;\langle\gamma,\alpha\rangle$}\\
             & BT for CCZ & $\llbracket81,3,(15,5)\rrbracket$ & $6/4$ & $2/4$ & 4 & {$(6,4,8)$} & --\\
            & BB init. & $\llbracket108,1,1;(9,9)\rrbracket$ & $4/4$ & $2/2$ & -- & {$(4,4,4)$} & $\langle\alpha\rangle;\langle\beta\rangle;\langle\gamma\rangle$\\
            & Escape BB & $\llbracket162,2,(18,18)\rrbracket$ & $8/8$ & $4/4$ & -- & $(8,8,8)$ & {$\langle\beta,\gamma\rangle;\langle\gamma,\alpha\rangle;\langle\alpha,\beta\rangle$}\\
            \hline
\end{tabular}
    \end{ruledtabular}
    \caption{\label{tab:codes}\textbf{BT/BB code parameters for CCZ cultivation.}
Each factory is labelled by the BT block size $n_{\mathrm{3D}}$ and the target fault distance $d_f$. BT and escape codes use $\llbracket n,k,(d_X,d_Z)\rrbracket$; initialization BB codes use subsystem parameters $\llbracket n,k,r;(d_X,d_Z)\rrbracket$, with $k$ protected and $r$ gauge logical qubits and dressed distances. $w_X,w_Z$ are check weights, $q_X,q_Z$ are qubit degrees, and $d_m$ is the meta-$Z$ distance. The tuple $(D_X,D_Z,D_{XZ})$ gives single-round CNOT depths for $X$-only, $Z$-only, and joint extraction. The SE circuits for the BT codes are optimized for the $Z$ fault distance rather than the CNOT depth. The logical image is the image of the composed map on $H_1$. For $n_{\mathrm{3D}}=21,36$, the escape rows list the image after fixing the initialized logical-$Z$ operator in each block. The defining polynomials and projection maps are given in \appref{App:CodeDetails} and \tabref{tab:CodeDetails}. The BT codes are from Refs.~\cite{li2026transversaldimensionjumpproduct, jacob2026singleshotdecodingfaulttolerantgates}.
}
\end{table*}

\subsection{Lifted BB codes and logical coupling}\label{Subsec:liftBB}
We consider projection-preserving lifts~\cite{hirasaki2026liftingliftedproductcodes} of the BB codes $Q_\bullet^\mathrm{AB}$, $Q_\bullet^\mathrm{BC}$, and $Q_\bullet^\mathrm{CA}$.
Each lift induces a chain map between the lifted code and the base code:
\begin{equation}\label{Eq:LiftedBBcoupling}
    \begin{tikzcd}
        {Q_2^{\mathrm{AB}}} & {Q_1^{\mathrm{AB}}} & {Q_0^{\mathrm{AB}}}\\
        {\widetilde Q_2^{\mathrm{AB}}} & {\widetilde Q_1^{\mathrm{AB}}} & {\widetilde Q_0^{\mathrm{AB}}}
        \arrow["p_2",from=2-1,to=1-1]
        \arrow["p_1",from=2-2,to=1-2]
        \arrow["p_0",from=2-3,to=1-3]
        \arrow[from=2-1,to=2-2]\arrow[from=2-2,to=2-3]
        \arrow[from=1-1,to=1-2]\arrow[from=1-2,to=1-3]
    \end{tikzcd}.
\end{equation}
We compose this map with the inclusion map $\iota$ in Eq.~\eqref{Eq:Inclusion} to couple the lifted BB code to the BT code:
\begin{equation}\label{Eq:LiftedBBBTcoupling}
    \begin{tikzcd}[column sep=1.4em,row sep=1.6em]
        {Q_3^{\mathrm{ABC}}} & {Q_2^{\mathrm{ABC}}} & {Q_1^{\mathrm{ABC}}} & {Q_0^{\mathrm{ABC}}}\\
        {} & {Q_2^{\mathrm{AB}}} & {Q_1^{\mathrm{AB}}} & {Q_0^{\mathrm{AB}}}\\
        {} & {\widetilde Q_2^{\mathrm{AB}}} & {\widetilde Q_1^{\mathrm{AB}}} & {\widetilde Q_0^{\mathrm{AB}}}
        \arrow["\partial_3",from=1-1,to=1-2]
        \arrow["\partial_2",from=1-2,to=1-3]
        \arrow["\partial_1",from=1-3,to=1-4]
        \arrow[from=2-2,to=2-3]\arrow[from=2-3,to=2-4]
        \arrow[from=3-2,to=3-3]\arrow[from=3-3,to=3-4]
        \arrow["\iota_2",from=2-2,to=1-2]
        \arrow["\iota_1",from=2-3,to=1-3]
        \arrow["\iota_0",from=2-4,to=1-4]
        \arrow["p_2",from=3-2,to=2-2]
        \arrow["p_1",from=3-3,to=2-3]
        \arrow["p_0",from=3-4,to=2-4]
    \end{tikzcd}.
\end{equation}
See \figref{Fig:LogicalCoupling}(a) for an illustration of the physical coupling. 

Index-2 lifts allow us to engineer the logical coupling as $\hat{p}_1$ need not be surjective and achieve selective initialization in the logical $Z$ basis through homomorphic measurement. Index-3 lifts preserve the transversal coupling of the logical qubits, allowing logical states in the BT codes to be faithfully transferred to high-distance lifted BB codes. See \figref{Fig:LogicalCoupling}(b) for an illustration of the logical coupling. Below, we describe BB-code lifting, selective initialization, and escape to lifted BB codes in more detail.

\vspace{5pt}
\noindent
\textbf{Projection-preserving lift.}
We consider a projection-preserving lift~\cite{hirasaki2026liftingliftedproductcodes} of $Q_\bullet^\mathrm{AB}$. Recall that BB codes are defined over the group algebra $R=\F_2[\Z_l\times \Z_m]$. We first extend the group $G = \Z_l\times \Z_m$ to $\widetilde{G} = \Z_l\times \Z_{\nu m}$ by $\Z_\nu$:
\begin{align}
    1\rightarrow \Z_\nu\rightarrow \Z_l\times \Z_{\nu m}\xrightarrow{\pi} \Z_l\times \Z_m.
\end{align}
The homomorphism $\pi$ induces the ring homomorphism
\begin{align}
 \Pi:\F_2[\widetilde G]&\longrightarrow\F_2[G],\nonumber\\
 \Pi \Big(\sum_{u,v}c_{uv}x^uy^v \Big)
 &=\sum_{u,v}c_{uv}x^uy^{v\bmod m}.
\end{align}
Choose lifted boundary polynomials $\widetilde a,\widetilde b\in\F_2[\widetilde G]$ satisfying $\Pi(\widetilde a)=a$ and $\Pi(\widetilde b)=b$. They define the lifted classical complexes $\widetilde A_\bullet$ and $\widetilde B_\bullet$ corresponding to Eq.~\eqref{Eq:BaseClassicalChain}. Their tensor product over $\F_2[\widetilde G]$ yields the lifted BB code $\widetilde{Q}^\mathrm{AB}_\bullet$. The degreewise projections
\begin{align}
 p_2=\Pi,\qquad p_1=\Pi\oplus\Pi,\qquad p_0=\Pi
\end{align}
define the chain map between $Q_\bullet^\mathrm{AB}$ and $\widetilde{Q}_\bullet^\mathrm{AB}$ in Eq.~\eqref{Eq:LiftedBBcoupling}.

\vspace{5pt}
\noindent
\textbf{Selective initialization.}
To measure $\overline{Z}_\alpha$ selectively, we couple the BT block to a lifted BB ancilla $\widetilde{Q}_\bullet^\mathrm{AB}$ through a homomorphic CNOT and read out the ancilla in the $Z$ basis.
The inclusion induces an injective map $\hat{\iota}_1$ on
logical-$Z$ classes, and the measured logical-$Z$ subspace satisfies
\begin{align}
    \operatorname{im}(\hat{\iota}_1\circ\hat{p}_1)
    \subseteq \operatorname{im}\hat{\iota}_1
    = \langle\overline{Z}_\alpha,\overline{Z}_\beta\rangle.
\end{align}
Measuring only $\overline{Z}_\alpha$ therefore requires choosing the lift so that the composite map has image $\langle\overline{Z}_\alpha\rangle$. To achieve this, we consider an even-index lift, for which the induced homology map $\hat{p}_1$ need not be surjective. Therefore, we construct an index-2 lifted BB code whose composite logical image is $\langle\overline{Z}_\alpha\rangle$, so the homomorphic measurement extracts only $\overline{Z}_\alpha$, leaving the other BT logical qubits unmeasured [\figref{Fig:LogicalCoupling}(b)]. Analogous index-two lifts enable selective measurements of $\overline{Z}_\beta$ and $\overline{Z}_\gamma$.

\vspace{5pt}
\noindent
\textbf{Escape BB codes.}  
At the escape stage, both retained BT logical qubits must couple faithfully to the logical qubits in the lifted BB memory codes. The projection $\hat p_1$ must therefore cover the full two-dimensional logical space of the BB codes. For an odd-index lift, the fibre-sum transfer $\hat\tau_1$ obeys $\hat p_1\hat\tau_1=I$ over $\F_2$, making the projection surjective and the transfer injective~\cite{hirasaki2026liftingliftedproductcodes}. Consequently, the two memory logical qubits are faithfully coupled to two BT logical qubits identified by the image $\operatorname{im}({\hat{\iota}_1}\circ\hat p_1)$ of the composite ${\hat{\iota}_1}\circ\hat p_1$. See \figref{Fig:LogicalCoupling}(b) for illustration.

Unlike weight-preserving lifts~\cite{guemard2024liftsquantumcsscodes, guémard2025moderatelengthliftedquantumtanner, symons2026sequencesbivariatebicyclecodes, guemard2025liftingcsscodehandlebody}, our construction allows the check weight to increase in the lifted BB codes used for escape. Indeed, weight-preserving lifts do not yield sufficient distance to store the cultivated CCZ states: a similar observation about the tradeoff between weight and distance is made in Refs.~\cite{wang2026checkweightconstrainedquantumcodesbounds, lee2026designprinciplesultrahighratequantum, lee2026designinggroupvaluedcodesregular}. The escape BB codes used in this study have check weight 8, while the base BB codes have check weight 4, as shown in \tabref{tab:codes}. For each logical qubit pair $\langle\alpha, \beta\rangle$, $\langle\beta, \gamma\rangle$, and $\langle\gamma, \alpha\rangle$, we lift the BB codes $Q_\bullet^\mathrm{AB}$, $Q_\bullet^\mathrm{BC}$, and $Q_\bullet^\mathrm{CA}$ to obtain lifted BB codes with check weight 8 and most of them achieve the distance upper bound $3d_\mathrm{base}$, where $d_\mathrm{base}$ is the base code distance.

\begin{figure}
    \centering
    \includegraphics[width=0.95\linewidth]{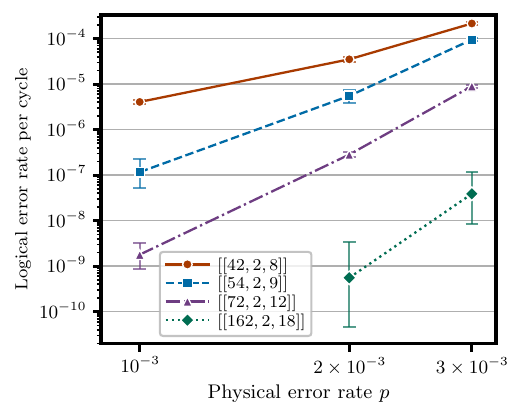}
    \vspace{-10pt}
    \caption{\textbf{Escape code memory error rate.} We perform memory experiments under our noise model in both the $X$-basis and the $Z$-basis. The logical error rate per cycle averaged over the two bases is shown with the error bars for the four BB codes used for escape at $p = 0.1\%, 0.2\%, 0.3\%$. The logical error rate for $\llbracket162, 2, 18\rrbracket$ at $p = 0.1\%$ is too small to estimate by sampling and is therefore omitted from the plot.}
    \label{fig:MemoryLER}
\end{figure}

To assess the escape codes as memories for the generated CCZ states, we perform memory experiments in both the $X$ and $Z$ bases using the telescoping decoder~\cite{bhardwaj2026highrateqldpcprocessors} except for the final mixed-integer programming (MIP) stage; the shots reaching the MIP stage are counted as logical failures. Figure~\ref{fig:MemoryLER} reports the block logical error rates per SE cycle at physical error rates $p=10^{-3}, 2\times10^{-3}$, and $3\times10^{-3}$. Since the cultivated resource spans three blocks, its per-cycle logical-fault probability is bounded by three times the per-block probability. Our results show that the memory error rate is small enough to store the cultivated CCZ states. We note that performing MIP would further reduce the logical error rate.

\section{Exact simulation}
We simulate the noisy CCZ cultivation factory using the error-propagation method of Ref.~\cite{fby6-xjbm}. Rather than simulating the non-Clifford gates directly, we propagate sampled circuit-level Pauli faults through the native CCZ circuit.  The CCZ gates transform these faults into Clifford errors acting on the ideal prepared state.  The ideal target $\ket{\overline{\mathrm{CCZ}}}^{\otimes 2}$ admits a four-term stabilizer decomposition, which remains a four-term decomposition under the propagated Clifford errors.  Subsequent Clifford operations and Pauli projections do not increase the number of components, allowing each trajectory to be simulated without truncating the coherent superposition.

We further accelerate mid-circuit measurements and adaptive feedback by exploiting the stabilizers shared by these components.  When calculating measurement probabilities, we retain interference between the components rather than treating them as a statistical mixture. For a $\chi$-term decomposition on $n$ qubits, direct evaluation of each Pauli expectation requires up to $\chi^2$ stabilizer overlaps, each computable in $O(n^3)$ time; repeating this calculation for $m$ measurements gives an $O(m\chi^2n^3)$ overlap-evaluation cost~\cite{bravyi2019simulationquantumcircuitslowrank, fby6-xjbm}. In our factory, the four components share $n-6$ independent stabilizers.  Stabilizer-tableau elimination separates this common stabilizer sector from a residual subsystem of at most six qubits.  We sample conditional Born probabilities and update the projected state using tableau operations and at most $64$ complex amplitudes, retaining interference and correlations between measurement outcomes. This avoids repeated full-system overlap evaluations; see \appref{app}.

To evaluate the output fidelity, we append a fictitious noiseless SE round and apply recovery.  We then compute the squared overlap of the recovered state with the ideal output and average over accepted attempts. The details of decoding and recovery are given in \appref{App:CorrectionRejection}.

\section{Conclusion}
We have developed an end-to-end CCZ-state cultivation factory by engineering the logical coupling between BT codes and lifted BB codes. Index-two lifts enable selective logical-$Z$ initialization, converting native hypergraph-state preparation into the direct production of two disjoint CCZ states. Index-three lifts preserve the coupling to the retained logical qubits while providing higher-distance BB memories for escape. Meta-checks, transversal readout, and decoder-based postselection detect faults throughout preparation and escape.

Exact simulations and resource accounting show a lower estimated spacetime volume per accepted CCZ state for our factory than for the surface-code $T$-cultivation-and-conversion benchmark at the corresponding output-infidelity target. BT codes with larger distances and low-depth cup-product gates could enable CCZ factories with higher fault distances~\cite{jacob2026singleshotdecodingfaulttolerantgates}. Although we focus on small BT/BB codes, our escape mechanism extends to magic-state preparation on other qLDPC codes using cup-product gates~\cite{breuckmann2024cups,breuckmann2026cupsgatesiioperads,li2026transversalnoncliffordgatesalmostgood,golowich2026improvedtransversalnoncliffordgates,li2026parallelizableaddressabletransversalnonclifford}, other constant-depth non-Clifford logical gates~\cite{golowich2024quantumldpccodestransversal}, or measurements of Clifford operators~\cite{christos2026nonabelianquantumlowdensityparity, zhu2026nonabelianqldpctqftformalism, chang2026magicquantumcodesurgery, williamson2026fastmagicstatepreparation, li2026nonabeliansheafquantumldpc}. While low acceptance remains a major source of overhead in our factory, higher physical-gate fidelity is expected to significantly reduce this cost by increasing acceptance, possibly achieving sufficient fidelity for factoring 2048-bit RSA integers~\cite{Zhou_2025}.

Our analysis of the observed failures suggests that escape-stage decoding currently limits the output fidelity rather than undetected pre-escape logical-$Z$ errors. This motivates jointly optimizing decoding and postselection to improve the tradeoff between accepted-output infidelity and rejection cost. More powerful decoders~\cite{beni2025tesseractsearchbaseddecoderquantum,leverrier2026approximatingoptimaldecodingquantum} and postselection methods~\cite{wills2026forcedgappostselectionquantum} may further improve this tradeoff at the cost of additional classical computation. CCZ-to-CCZ distillation~\cite{gong2026magicstatedistillationcodes} offers a complementary route to suppressing residual errors, provided correlations between cultivated outputs are taken into account. By generating CCZ states directly, our approach avoids $T$-state conversion and offers a promising route to compact factories with low spacetime overhead.

\section*{Acknowledgements}
We thank Christopher A. Pattison for suggesting a potential connection between Ref.~\cite{hirasaki2026liftingliftedproductcodes} and magic state cultivation during the KIAS workshop ``Quantum Information and Dynamics 2026: qLDPC Theory and Application''. We also thank Isaac Kim for sharing the code for exact magic state simulation~\cite{fby6-xjbm, kim2026xtim}. 
YH thanks Zejun Liu for fruitful discussions.  YH is supported by a Japan Student Services Organization (JASSO) scholarship. JYL is supported by a faculty startup grant at the University of Illinois, Urbana-Champaign. Computations are performed using resources provided by the Illinois Campus Cluster Program at the University of Illinois, Urbana-Champaign, and services provided by the OSG Consortium~\cite{Pordes2007OpenScienceGrid, Sfiligoi2009GlideinWMS, OSG2006OSPool, OSG2015OSDF}, which is supported by National Science Foundation awards \#2030508 and \#2323298. YH and JYL acknowledge support from KIAS through the Quantum Universe Center scholar program. The authors acknowledge the use of OpenAI Codex to generate numerical simulation codes and assist with writing the manuscript. The idea of the lifting-based CCZ cultivation factory was conceived and developed by the authors.

\section*{Data and code availability}
The simulation code and accompanying benchmark inputs are publicly available~\cite{hirasaki2026cczfactorycode}.

\appendix

\section{Factory details}\label{App:cultivationdetails}
\begin{figure*}
    \centering
    \includegraphics[width=1.0\linewidth]{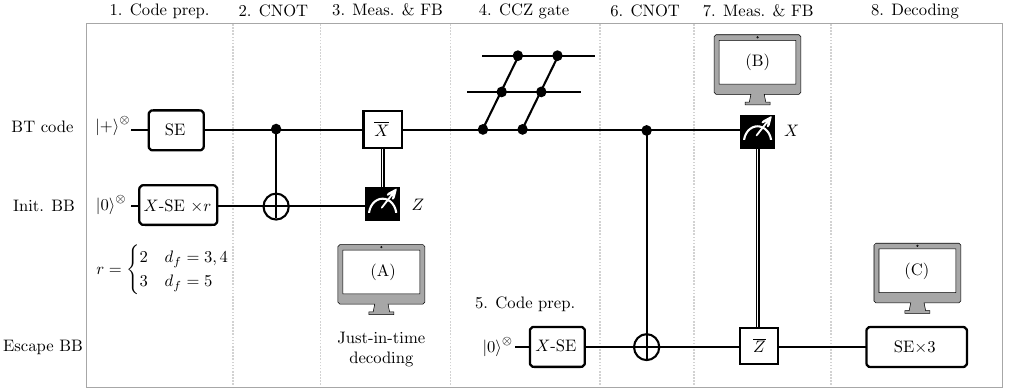}
    \caption{\textbf{CCZ factory pipeline.} {\bf 1.} The BT and BB codes are initialized in the logical $+$ and $0$ states, respectively. {\bf 2.} A homomorphic CNOT is applied with physical depth two. {\bf 3.} The BB code is measured transversally in the $Z$ basis; the outcomes are decoded, and adaptive feedback is applied. {\bf 4.} The cup-product CCZ gate is applied to the three separately prepared BT code blocks. {\bf 5.} The escape BB code is prepared in the logical $0$ state. {\bf 6.} A homomorphic CNOT is applied with physical depth three. {\bf 7.} The BT code block is measured transversally in the $X$ basis; the outcomes are decoded, and feedback is applied. {\bf 8.} Three rounds of SE are performed; the outcomes are decoded, and postselection is performed based on the decoder output.}\label{Fig:pipeline}
\end{figure*}

We describe each stage of the cultivation factory below. Figure~\ref{Fig:pipeline} shows the complete pipeline, Fig.~\ref{Fig:jointdecoding} shows how measurement outcomes from different blocks are jointly decoded, and Fig.~\ref{Fig:faultmap} illustrates faults that can produce multiple errors and how they are detected.
We first describe the physical operations, measurements, and feedback at each stage of the pipeline in \appref{App:Pipeline}. We then identify faults that can propagate to multiple errors or cause incorrect feedback, and explain how they are detected in \appref{App:DangerousFault}. Finally, we describe the joint decoding and postselection rules used to correct errors and reject unreliable attempts in \appref{App:CorrectionRejection}.

\subsection{Pipeline}\label{App:Pipeline}
\vspace{5pt}
\noindent
\textbf{1. Code preparation.}
We prepare each BT block in $\ket{\overline{+++}}$ by initializing all physical qubits in $\ket{+}$ and measuring the $Z$ checks once. Even in the absence of noise, the outcomes are random, but an outcome-dependent Pauli-frame update fixes the $Z$ checks to $+1$ while preserving the logical $+$ states. In the presence of noise, the measurement outcomes may not reflect the true $Z$-check values, which can lead to unbounded errors at later stages. We use the BT meta-$Z$ constraints to detect and correct such measurement errors; details are given in \appref{App:DangerousFault}.

Additionally, we measure the BT $X$ checks once. Nontrivial $X$-check outcomes indicate the presence of local $Z$ errors in the BT code or errors in the measurements themselves. We reject any attempt with nontrivial $X$-check outcomes. Empirically, this postselection reduces the output infidelity by a constant factor by detecting $Z$ errors introduced during the $Z$-check measurements. We note that hook errors from the $Z$-check measurements do not reduce the $Z$ fault distance, so omitting the additional $X$-check measurements increases the output infidelity only by a constant factor.

We prepare the BB codes in the logical $0$ state by initializing all physical qubits in $\ket{0}$ and measuring the $X$ checks. These measurement outcomes are also random. As with the BT codes, incorrectly inferred $X$-check values can lead to logical errors at later stages. Since the BB codes do not possess meta-checks that allow correction, we use repetition checks and postselection. We measure the $X$ checks $r$ times and postselect on the resulting records, where $r = 2$ for $d_f = 3, 4$ and $r = 3$ for $d_f = 5$; details are given in \appref{App:DangerousFault}.

\vspace{5pt}
\noindent
\textbf{2. Homomorphic CNOT.}
We apply a homomorphic CNOT from the BT block to the lifted initialization BB block. The coupling is specified by the chain map $f=\iota\circ p$, which composes the projection and inclusion in \eqnref{Eq:LiftedBBBTcoupling}. Its qubit component $f_1$ specifies a physical CNOT from BT qubit $i$ to BB qubit $j$ whenever $(f_1)_{ij}=1$:
\begin{equation}
    \begin{tikzcd}
        Q_3 & {Q_2} & {Q_1} & {Q_0}\\
        {} & {\widetilde Q_2}^{\mathrm{init.}} & {\widetilde Q_1}^\mathrm{init.} & {\widetilde Q_0}^\mathrm{init.}
        \arrow["\partial_3", from=1-1, to=1-2]
        \arrow["\partial_2", from=1-2, to=1-3]
        \arrow["\partial_1", from=1-3, to=1-4]
        \arrow["\widetilde\partial_2", from=2-2, to=2-3]\arrow["\widetilde\partial_1", from=2-3, to=2-4]
        \arrow["f_2", from=2-2, to=1-2]
        \arrow["f_1", from=2-3, to=1-3]
        \arrow["f_0", from=2-4, to=1-4]
    \end{tikzcd}.
\end{equation}
Under the index-two projection, each base BB qubit has two lifted preimages, and the inclusion maps distinct base BB qubits to distinct BT qubits. Thus, each participating BT qubit controls two CNOTs, while each lifted BB qubit is targeted once. Assigning the two CNOTs from each BT qubit to different layers gives a depth-two circuit with disjoint gates within each layer. Because each BB qubit couples to only one BT qubit, a single BB $Z$ error can propagate to at most one BT qubit.

The chain-map identities determine the action on the stabilizer generators:
\begin{align}\label{Eq:stabTransformX}
    H_{X, i}^\mathrm{BT}\mapsto H_{X, i}^\mathrm{BT}\prod_{({f}_0)_{ij} = 1}H_{X, j}^\mathrm{BB},\\
    \label{Eq:stabTransformZ}
    H_{Z, j}^\mathrm{BB}\mapsto H_{Z, j}^\mathrm{BB}\prod_{({f}_2)_{ij} = 1}H_{Z, i}^\mathrm{BT}.
\end{align}
The BT $Z$-check generators and BB $X$-check generators remain unchanged. These transformations couple the check values of the two blocks, so their measurement records must be decoded jointly, as shown in Fig.~\ref{Fig:jointdecoding}. The decoding procedure is described in \appref{App:CorrectionRejection}.

\vspace{5pt}
\noindent
\textbf{3. Transversal measurements and feedback.}
The physical qubits in the lifted BB code are transversally measured in the $Z$-basis. This yields the $Z$-check values and logical-$Z$ outcomes. $Z$-syndromes are calculated together with the (potentially corrected) $Z$-check outcomes of the BT code based on Eq.~\eqref{Eq:stabTransformZ}. This joint decoding is also shown in \figref{Fig:jointdecoding}(A).

To improve the overall performance, we correct errors at this stage with partial postselection. The decoding strategy and postselection rule are elaborated in \appref{App:CorrectionRejection}. Using the decoded logical-$Z$ outcomes, we apply conditional logical-$X$ corrections to prepare the three BT blocks in $\ket{\overline{0++}}$, $\ket{\overline{+0+}}$, and $\ket{\overline{++0}}$, respectively. We also apply physical $X$ corrections to set the inferred BT $Z$-check signs to $+1$. In this protocol, all feedback is implemented physically before the CCZ layer. Deferring these corrections would generally require a Clifford frame, since conjugating an $X$ correction through a CCZ gate introduces a CZ component. Related just-in-time decoding for logical magic-state preparation and non-Clifford circuits is discussed in Refs.~\cite{bauer2026planarfaulttolerantcircuitsnoncliffo,vrty-qs5h, bombin20182dquantumcomputation3d, Ruiz2026, Brown2020}.

\vspace{5pt}
\noindent
\textbf{4. Cup-product CCZ gate.}
We apply the depth-2 CCZ gate $U_\cup$ in Eq.~\eqref{eq:CCZ_cup} across the three BT code blocks. The three BT codes are initialized in the $\ket{\overline{0++}}$, $\ket{\overline{+0+}}$, and $\ket{\overline{++0}}$ states, respectively. After applying the CCZ gates, the resulting logical state is
\begin{align}
    \ket{\overline{0}}^{\otimes 3}\otimes \ket{\overline{\mathrm{CCZ}}}^{\otimes 2}.
\end{align}

\vspace{5pt}
\noindent
\textbf{5. Escape code preparation.}
The escape BB code is prepared in the logical $\ket{\overline{00}}$ state by performing one round of $X$-SE on qubits initialized in $\ket{0}$ states. Measurement errors at this stage lead to incorrect decoder inference of (B) as shown in \figref{Fig:jointdecoding}, and the measurement outcomes should be re-decoded together with the subsequent three rounds of SE: the details are provided in \appref{App:CorrectionRejection}.

\vspace{5pt}
\noindent
\textbf{6. Homomorphic CNOT. }
We perform the homomorphic CNOT between the BT code and the escape BB code for the logical state transfer. We note that the stabilizers of the BT code and the escape BB code are mixed in a similar manner in Eqs.~\eqref{Eq:stabTransformX} and \eqref{Eq:stabTransformZ}. The corresponding joint decoding is shown in Fig.~\ref{Fig:jointdecoding}. Here we use the \emph{deformed chain maps} discussed in \appref{App:chainmapDeformation}. The resulting homomorphic CNOT has depth three and couples every BT physical qubit to BB physical qubits, enabling the postselection illustrated in \figref{Fig:faultmap}. The details of the postselection are discussed in \appref{App:DangerousFault}.

\vspace{5pt}
\noindent
\textbf{7. Transversal measurement and feedback.}
The physical qubits in the BT code are transversally measured in the $X$-basis. This yields information about the $X$ checks and logical-$X$ values. Again, the $X$-check values in the BT code are inferred jointly with the $X$-check values of the escape BB codes as shown in \figref{Fig:jointdecoding}. The BB $X$-check values at this stage are noisy and require correction using subsequent rounds of SE. To reduce the spacetime volume incurred by rejection at a later stage, we perform early postselection here.

If the global meta-$X$ check does not detect measurement errors on the escape BB code, we decode the $X$-checks on the BT codes and perform postselection for non-trivial $X$-check outcomes. If the global meta-$X$ check signals measurement errors, we defer the rejection to the decoding stage. This allows us to retain shots that would otherwise be rejected solely because the BB $X$ checks were measured incorrectly. These $X$-check outcomes are then jointly decoded again to identify $X$-syndromes hidden by the incorrect measurement of BB code $X$-checks as shown in \figref{Fig:jointdecoding}.

\vspace{5pt}
\noindent
\textbf{8. Decoding and postselection.}
We perform additional SE rounds for postselection. The outcomes are decoded and we perform postselection based on the decoder output. The detailed strategy is elaborated in \appref{App:CorrectionRejection}.

\subsection{Dangerous faults and their detection}\label{App:DangerousFault}
\begin{figure*}
    \centering
    \includegraphics[width=1.0\linewidth]{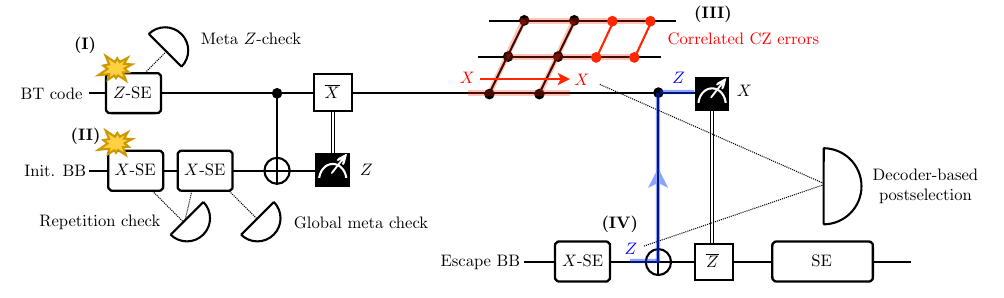}
    \caption{\textbf{Fault location and detection.} We show the faults that can lead to multiple errors in the factory, and describe how these dangerous faults are detected. Some circuit elements are omitted for simplicity. }\label{Fig:faultmap}
\end{figure*}

For each factory, we target a fault distance equal to the $Z$ distance of the BT code by rejecting attempts with nontrivial BT $X$ syndromes. However, correlated errors and measurement errors that cause incorrect decoder inference can reduce the fault distance, requiring additional postselection. Figure~\ref{Fig:faultmap} summarizes these dangerous faults and how they are detected. Below, we explain each fault mechanism and how they can reduce the fault distance; the detailed procedures for error correction and postselection are given in \appref{App:CorrectionRejection}.

\begin{figure}
    \centering
    \includegraphics[width=1.0\linewidth]{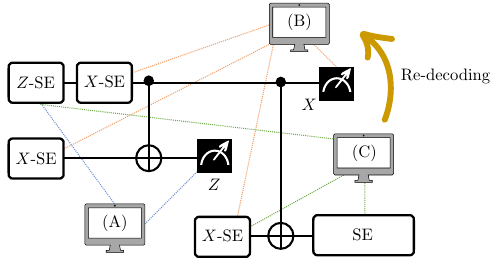}
    \caption{\textbf{Joint decoding.} Homomorphic CNOTs mix the stabilizer values of the BT, initialization BB, and escape BB codes, so decoding should be performed jointly. The dotted lines show which measurement outcomes are jointly decoded. The outcomes supplied to decoder (B) are re-decoded using the output of decoder (C), as indicated by the yellow arrow.}\label{Fig:jointdecoding}
\end{figure}

\vspace{5pt}
\noindent
\textbf{(I) BT $Z$-check measurement errors.}
In the presence of noise, the measured outcomes can differ from the actual stabilizer values, leading to incorrect inference by decoder (A) in Fig.~\ref{Fig:jointdecoding}. Such inference can produce incorrect $X$ feedback when setting the BT $Z$ checks to $+1$ before the CCZ layer. The resulting residual $X$ errors can propagate through the CCZ gates into multiple CZ errors on the other BT blocks, complicating recovery. We therefore use the meta-$Z$ check matrix $M_Z$ defined in \eqnref{eq:3D_complex} to detect and, when possible, correct measurement errors.

Let $s_Z$ denote the binary vector of measured $Z$-check outcomes, with meta-check syndrome $m_Z=M_Zs_Z$. When $m_Z\neq 0$, we use the Relay-BP decoder~\cite{muller2025relay} to infer a measurement-error pattern and flip the corresponding bits of $s_Z$~\cite{PhysRevX.5.031043, PRXQuantum.2.020340}. However, the meta-check distance in \tabref{tab:codes} is insufficient to preserve the target factory fault distance through correction alone. To retain acceptance while limiting miscorrections, we allow corrections of Hamming weight at most one and reject corrections with a larger weight. This early rejection avoids spending resources on later stages of those attempts. Remaining miscorrections are checked using the additional syndrome information available to joint decoders (A) and (C) in Fig.~\ref{Fig:jointdecoding}; attempts inconsistent with the earlier repair are rejected. The detailed consistency checks are described in \appref{App:CorrectionRejection} (C).

\vspace{5pt}
\noindent
\textbf{(II) Init. BB $X$-check measurement errors.}
Although the initialization BB $X$-check outcomes are random even without noise, reliable outcomes allow a Pauli-frame update that sets all stabilizer signs to $+1$ and prepares the logical $0$ state. Measurement errors on BB $X$-checks can instead cause decoder (B) in Fig.~\ref{Fig:jointdecoding} to infer incorrect BT $X$-check values, potentially reducing the factory fault distance. We therefore postselect on the initialization BB $X$-check records so that their detection distance against measurement errors is at least the target factory fault distance $d_f$.

The initialization BB codes satisfy the global meta-check $\prod_i H_{X,i}^{\mathrm{BB}}=1$. We require a trivial meta-check syndrome in every round and agreement of the $X$-check records across all rounds. We use two rounds for $d_f=3,4$ and three for $d_f=5$. Together, these conditions give detection distances of four and six, respectively, against measurement-bit errors.

\vspace{5pt}
\noindent
\textbf{(III) Correlated CZ errors by CCZ gates.}
The depth-2 cup-product gate $U_\cup$ can propagate an $X$ error on one block to correlated CZ errors on the other two blocks, as shown in \figref{Fig:faultmap}. CCZ gates propagate a physical $X$ error according to
\begin{align}\label{Eq:CCZ_XCZ}
    X_1\mapsto X_1CZ_{23} = \frac{1}{2}X_1(I_{23} + Z_2 + Z_3 - Z_2Z_3)
\end{align}
Crucially, the depth-two CCZ circuit can create two $Z$ errors in one BT code block, which could reduce the fault distance of the protocol. These correlated errors can be rejected by detecting correlated $X$ errors on the BB code block that signal the presence of $X$ errors on the BT code block, as illustrated in \figref{Fig:faultmap} and discussed below for the decoding stage.

\vspace{5pt}
\noindent
\textbf{(IV) Multiple $Z$ error back propagation.}
The escape homomorphic CNOTs use the deformed chain maps described in \appref{App:chainmapDeformation} and \figref{Fig:CMdeformation}, which couple certain BB qubits to multiple BT qubits. A single $Z$ fault on a BB qubit can therefore propagate backward through these CNOTs to multiple BT qubits, potentially reducing the factory fault distance. Subsequent BB $X$-check measurements provide syndrome information about these $Z$ faults. We use this information for postselection to reject the faults to leading order and restore the target fault distance.

At both noise strengths, we reject an attempt if the detection-event pattern from the first noisy BB $X$-check round contains the full syndrome of a single BB $Z$ fault that can occur before or between the homomorphic CNOT layers and propagate to at least two BT qubits. At $p=0.01\%$, we also check each completed escape-decoder recovery. We reject the attempt if any selected mechanism admits a compatible BB or ancilla fault that propagates $Z$ errors to at least two qubits in the same BT block.

\subsection{Error correction and postselection}\label{App:CorrectionRejection}
We explain the error correction strategy and how postselection is performed at each decoding stage.

\vspace{5pt}
\noindent
\textbf{(A) Transversal $Z$ measurement.}
Transversal $Z$ measurements of the initialization BB blocks yield both their $Z$-check values and logical-$Z$ outcomes. By \eqnref{Eq:stabTransformZ}, each measured BB $Z$-check value depends on the corresponding pre-CNOT BB check and the BT checks coupled to it. We therefore combine these outcomes with the earlier BT $Z$-check records and decode the two codes jointly under the circuit-level noise model.

The BT $Z$-check records may still contain measurement errors after the meta-check correction. We use the records repaired by Relay-BP as the reference and include additional measurement-bit flips with probability $p$ in the joint decoding model~\cite{PRXQuantum.4.020332}. This allows the decoder to account for residual errors in the earlier records when inferring the correction.

We then apply a second decoding pass with argument reweighting~\cite{sv11-f78l}. For this pass, the probabilities of the physical error mechanisms selected by the first recovery are updated as $p_j\leftarrow p_j^{1.2}$. We accept the attempt only if the two passes agree on the inferred logical-$Z$ correction. On acceptance, we retain the first recovery for the physical feedback before the CCZ layer. To suppress logical initialization errors when the fault distance against $X$ errors is low, we cap the inferred fault weight, and reject the shot if the decoder output exceeds the cap. For $n_{\mathrm{3D}}=81$, we allow at most two inferred BB $Z$-readout errors per block on each completed initialization recovery. For $n_{\mathrm{3D}}=21,36$, we cap the combined weight of the retained joint recovery and the meta-check repair at $(3,1,1)$ and $(1,0,2)$, respectively, in block order.

\vspace{5pt}
\noindent
\textbf{(B) Transversal $X$ measurement.}
Similarly, transversal $X$ measurements of the BT codes yield both their $X$-check values and logical-$X$ outcomes. Because the BT codes have interacted with both the initialization and escape BB codes, inferring the BT syndrome requires the $X$-check records from both sets of BB codes, as shown in Fig.~\ref{Fig:jointdecoding}. Postselection on the initialization BB $X$-check records restores the required detection distance against measurement errors, as described in \appref{App:DangerousFault}. We therefore focus on errors in the escape BB records.

Repeated SE rounds provide additional information for correcting errors in the escape BB measurement records. However, waiting until these rounds are complete before rejecting an attempt incurs an additional spacetime cost. We therefore apply postselection at two stages: immediately after the transversal $X$ measurement and again after the repeated SE rounds.

At the early stage, a single escape-BB measurement error can produce a non-trivial BT syndrome and cause an $O(p)$ loss in acceptance through false rejection. We therefore first test the global meta-$X$ check. If it is trivial, we reject attempts with nontrivial inferred BT syndromes; otherwise, we defer the decision to decoder (C). This meta-check has measurement-error detection distance two, so some errors can remain undetected. Decoder (C) therefore rechecks every surviving attempt using the subsequent BB SE rounds to restore the target fault distance.

\vspace{5pt}
\noindent
\textbf{(C) Decoding escape BB code.} We perform $w_c+w_b$ noisy rounds of both $X$- and $Z$-SE, consisting of $w_c$ committed rounds followed by $w_b$ buffer rounds. We use $w_c=2$ and $w_b=1$ for all factories. The syndrome histories of the BT and BB codes are decoded jointly using the Telescoping decoder with the GARI transformation~\cite{Maan2026}. This stage consists of two decoding steps.

The first step checks the consistency of the earlier BT meta-check correction. We jointly decode the BT, initialization BB, and escape BB measurement records through the first round of escape-BB SE using an $X$-error detector error model (DEM). The model includes $X$ errors, noisy $Z$-check measurements, and the BT meta-$Z$ constraints~\cite{PRXQuantum.4.020332}. Because the homomorphic CNOTs encode the BT $Z$-check values in the BB check outcomes, the later measurements can reveal earlier miscorrections at leading order. Through the homomorphic CNOTs, the initialization and escape BB $Z$-check measurements jointly probe the full BT $Z$-stabilizer space; see \appref{App:chainmapDeformation} for the chain-map construction and \figref{Fig:CMdeformation}(c) for illustration. We reject an attempt if the inferred BT measurement-error correction disagrees with the repair used for physical feedback before the CCZ gate; otherwise, we fix the repaired BT check values and proceed to the second decoding step.

The second step uses a correlated circuit-level DEM of the BT--BB escape circuit to correct the remaining errors and inform postselection. The model includes BT $X$-readout noise, BB preparation, fanout gates, and noisy BB SE. Its detectors encode the BT $X$-check outcomes, BB preparation meta-checks, and BB $X/Z$ SE outcomes. Errors on the BT qubits before the homomorphic CNOTs are approximated by local stochastic Pauli errors, with probabilities accumulated from initialization, physical feedback, and the CCZ layers. In particular, we approximate the CZ errors generated by CCZ propagation using their Pauli-twirled marginals, assigning a conditional $Z$-error probability of one half to each affected target. This approximation decouples the three blocks and reduces decoding complexity. The Clifford escape circuits retain their circuit-level correlated error mechanisms. The DEM is therefore approximate, while the noisy factory simulation remains exact.

We apply additional postselection to the decoder output. First, we subtract the inferred contribution of escape-BB preparation errors from the BT $X$-check values and reject any nontrivial revised $X$ charge. We also reject inferred incoming BT $X/Y$ errors before the homomorphic CNOTs, which signal the correlated CZ errors discussed in \appref{App:DangerousFault}. Finally, we reject an attempt if an inferred BB $Z$ error is consistent with a physical fault that propagates backward through the homomorphic CNOTs to a logical $Z$ error on the BT code, leaving its $X$-check syndrome trivial. These postselection rules reject fault mechanisms that would otherwise reduce the factory fault distance.

We also use argument-reweighting postselection~\cite{sv11-f78l} to assess confidence in the decoded logical class. The depth-three homomorphic CNOTs can reduce the fault distance during escape relative to that of the BB memory, so we use additional postselection to suppress logical errors. This serves a similar purpose to complementary-gap tests~\cite{PRXQuantum.5.010302, Gidney2025, Lee2026, meister2024efficientsoftoutputdecoderssurface, PhysRevA.89.022326} in $T$-state cultivation~\cite{gidney2024magic,sahay2025fold}. We run the decoder up to three times on the same records from all $w_c+w_b$ noisy SE rounds including the initial decoding. Before each subsequent decoding, we cumulatively reweight the probabilities of the physical error mechanisms inferred by the previous decoding as $p_j\leftarrow p_j^{1.2}$, including mechanisms in the buffer rounds. GARI auxiliary variables are not reweighted. The argument-reweighting criterion requires all decoder outputs to agree on the logical class at the end of the $w_c$ committed rounds. In addition to this criterion, we apply the decoder-output postselection rules described above to every decoding round. Empirically, this combined postselection reduces the output infidelity by a constant factor.

\vspace{5pt}
\noindent
\textbf{Fictitious perfect SE and recovery.}
After postselection, we append a fictitious round of perfect SE to evaluate the output fidelity. We call the error nodes before the first noisy escape SE round the \emph{entrance stage}. We first fix its error mechanisms to the pattern inferred by the preceding noisy decoding. Without this constraint, the decoder often selects an incorrect entrance logical class, resulting in the logical error at the end. Fixing entrance recovery avoids such logical errors in the subsequent memory decoding. We run Telescoping with the entrance error mechanisms fixed and require its logical class at the end of the commitment window to agree with the previously accepted class. If it does not, it heralds logical errors so we turn to more accurate (but perhaps slower) decoders: we try Tesseract~\cite{beni2025tesseractsearchbaseddecoderquantum} and Telescoping for both fixed DEM, and Telescoping with unfixed DEM. Among recoveries that satisfy the logical-agreement condition, we choose the one with the lowest cost. If none satisfies the logical-agreement condition, we use MIP~\cite{bhardwaj2026highrateqldpcprocessors} on the unfixed DEM and retain the lowest-cost recovery found. No further postselection is performed during this fidelity evaluation.

In our simulations, fewer than a fraction $10^{-3}$ of the shots reach the MIP stage. Most of these cases are solved within a few minutes, while approximately $10^{-6}$ of the shots require six-hour runs. In practice, if the output states undergo further distillation~\cite{gong2026magicstatedistillationcodes}, additional postselection can be applied at injection into the distillation protocol, allowing shots with difficult syndromes to be discarded without interrupting the main computation.

\subsection{Chain map deformation}\label{App:chainmapDeformation}
\begin{figure}
    \centering
    \includegraphics[width=1.0\linewidth]{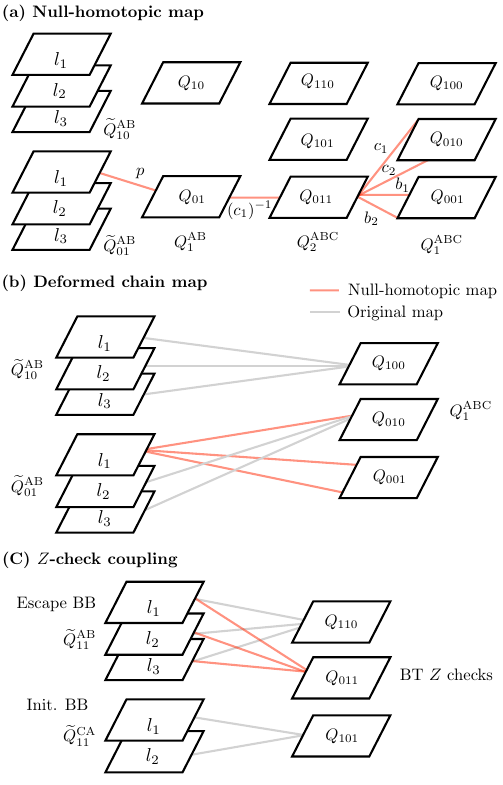}
    \caption{\textbf{Chain map rewiring.} We add the null-homotopic map in \textbf{(a)} to the original composite chain map in Eq.~\eqref{Eq:LiftedBBBTcoupling} to obtain the deformed map \textbf{(b)}. The labels $l_i$, $i=1,2,3$, specify the fibre coordinate: for each projected group element $g$, the pairs $(g,l_i)$ label its three lifted preimages. We use the shorthand notation $Q_{100} = A_1\otimes B_0\otimes C_0$. For the BT codes studied here, the boundary maps $a$, $b$, and $c$ in Eq.~\eqref{Eq:BaseClassicalChain} support two group elements, which we denote by $a_1, a_2$, $b_1, b_2$, and $c_1, c_2$, respectively. \textbf{(c)} $Z$-checks on the BT codes couple to $Z$-checks on the initialization/escape BB codes through homomorphic CNOTs. The grey lines describe the coupling through the original chain maps in Eq.~\eqref{Eq:LiftedBBBTcoupling}, and the red lines describe the coupling through the null-homotopic map $h_1\widetilde{\partial}_{2}^\mathrm{AB}$.}\label{Fig:CMdeformation}
\end{figure}

As shown in \figref{Fig:faultmap}, the $X$ errors on BT codes should be detected by the BB code checks through homomorphic CNOT propagation. However, the undeformed chain map in \eqnref{Eq:LiftedBBBTcoupling} couples two of the three BT physical sectors and leaves the third untouched. Thus, we deform the chain map by adding a null-homotopic map to couple every BT data qubit while preserving the induced logical map~\cite{sahay2026computingqldpccodesclimbing,berent2026liftedsurgeryfastprocessing,weibel1994introduction,hatcher2002algebraic}. Together with the initialization map, the deformed escape map also couples all three BT $Z$-check sectors to BB $Z$-checks, enabling the consistency test described in \appref{App:CorrectionRejection}; see \figref{Fig:CMdeformation}.

We describe the construction for the chain map $f_\bullet:\widetilde{Q}_\bullet^{\mathrm{AB}}\to Q_\bullet^{\mathrm{ABC}}$ from the escape BB code to the BT code, which is used for the third block of the $n_{\mathrm{3D}}=27$ factory. Let $f_i$ denote the components of the undeformed chain map. Given $\mathbb{F}_2$-linear maps $h_i:\widetilde{Q}_i^{\mathrm{AB}}\to Q_{i+1}^{\mathrm{ABC}}$, the null-homotopic deformation is
\begin{align}
    f_i\mapsto \widetilde{f}_i =  f_i + \partial_{i + 1}^{\mathrm{ABC}}h_i + h_{i - 1}\widetilde{\partial}_i^{\mathrm{AB}}.
\end{align}
This is schematically described by the following diagram:
\begin{equation}
    \begin{tikzcd}
         {Q_2^{\mathrm{ABC}}} & {Q_1^{\mathrm{ABC}}} & {Q_0^{\mathrm{ABC}}}\\
        {\widetilde{Q}_2^{\mathrm{AB}}} & {\widetilde{Q}_1^{\mathrm{AB}}} & {\widetilde{Q}_0^{\mathrm{AB}}}
        \arrow["\partial_2^\mathrm{ABC}", from=1-1, to=1-2]\arrow["\partial_1^\mathrm{ABC}", from=1-2, to=1-3]
        \arrow["\widetilde{\partial}_2^\mathrm{AB}", from=2-1, to=2-2]\arrow["\widetilde{\partial}_1^\mathrm{AB}", from=2-2, to=2-3]
        \arrow["f_2", from=2-1, to=1-1]
        \arrow["h_1"', from=2-2, to=1-1]
        \arrow["h_0"', from=2-3, to=1-2]
        \arrow["f_1", from=2-2, to=1-2]
        \arrow["f_0", from=2-3, to=1-3]
    \end{tikzcd}.
\end{equation}
The deformed chain maps have the same logical action on the homology groups.

We set $h_i=0$ for $i\neq1$ and choose $h_1$ so that every BT qubit couples to BB qubits while the circuit depth remains three. We label each qubit of $\widetilde{Q}_{01}^{\mathrm{AB}}$ by its projected group element $g$ and fibre coordinate $l_i$. We define $h_1(g,l_1)=c_1^{-1}g\in Q_{011}$, set $h_1$ to zero on all remaining qubits, and extend linearly; see \figref{Fig:CMdeformation}(a). The resulting qubit map is $\widetilde f_1=f_1+\partial_2^{\mathrm{ABC}}h_1$. For a qubit $(g,l_1)\in\widetilde{Q}_{01}^{\mathrm{AB}}$, the null-homotopic contribution couples it to $g$ and $gc_2c_1^{-1}$ in $Q_{010}$, and to $gb_1c_1^{-1}$ and $gb_2c_1^{-1}$ in $Q_{001}$. The coupling to $g\in Q_{010}$ cancels with the original chain-map coupling, leaving three couplings and preserving depth three; see \figref{Fig:CMdeformation}(b).

The corresponding $Z$-check map is $\widetilde f_2=f_2+h_1\widetilde{\partial}_2^{\mathrm{AB}}$. Figure~\ref{Fig:CMdeformation} illustrates the resulting $Z$-check couplings with both initialization and escape BB codes. We use the BB code $\widetilde{Q}_{\bullet}^\mathrm{CA}$ for initialization. The BT $Z$-checks in $Q_{101} (Q_{110})$ couple to the $Z$-checks of $\widetilde{Q}_{\bullet}^\mathrm{CA} (\widetilde{Q}_{\bullet}^\mathrm{AB})$ through the undeformed chain maps, and $Z$-checks in $Q_{011}$ couple to the $Z$ checks in $\widetilde{Q}_{\bullet}^\mathrm{AB}$ through the null-homotopic map. The same construction applies to the other blocks and factory sizes, with the maps specified in \tabref{tab:escape-homomorphisms}.

\section{Details of resource estimation}\label{App:ResourceEstimate}

We explain the calculation of the expected spacetime volume in \tabref{tab:factory-benchmark}. First we define the expected spacetime volume following Appendix E.3 of Ref.~\cite{sahay2025fold}, and elaborate on the CCZ state factory ($T$-state based factory) in \appref{app:spacetime-accounting} (\appref{App:TcultivationCCZ}).

Consider a circuit with $M$ postselection stages and $M+1$ intervening circuit segments $C_0,\ldots,C_M$. For a segment with $Q_i$ active qubits and $D_i$ gate steps (circuit depth), the spacetime volume is $V_i=Q_iD_i$. If the number of active qubits varies, we instead sum over gate steps, $V_i=\sum_t Q_{i,t}$, including data qubits, ancillas, and qubits retaining quantum states. Let $f_i$ be the fraction of shots that pass the first $i$ postselection stages, so that $f_0=1$ and $f_M$ is the acceptance probability for this circuit. The expected spacetime volume per successful shot is
\begin{align}
 V=\frac{1}{f_M}\sum_{i=0}^{M}f_iV_i.
 \label{eq:volume-accounting}
\end{align}
The weights $f_i$ count each segment only when it is reached, including the executed portions of rejected shots; division by $f_M$ accounts for repetitions until success. If no physical operations follow the final postselection stage, $V_M=0$. For resource accounting, we count a CNOT, an individual single-qubit gate, a $Z$-basis reset and a $Z$-basis measurement as one step each. We count $X$-basis preparation and readout as two steps each, including the Hadamard basis change, following Ref.~\cite{sahay2025fold}. The noisy simulations retain their circuit-level compilations.

\subsection{CCZ factory} \label{app:spacetime-accounting}

\begin{table*}[t]
\centering
\small
\renewcommand{\arraystretch}{1.25}
\setlength{\tabcolsep}{2.5pt}
\begin{ruledtabular}
\begin{tabular}{l|rrrr|rrrr}
& \multicolumn{4}{c|}{$(n_{\mathrm{3D}},d_f)=(21,3)$} & \multicolumn{4}{c}{$(27,3)$}\\
Rejection stage & $Q$ & $D$ & $f$ & $2\overline V_{\le i}$ & $Q$ & $D$ & $f$ & $2\overline V_{\le i}$\\
\hline
Init. BB preparation & 126 & 20 & $0.83$ & $3.02\times10^{3}$ & 162 & 16 & $0.84$ & $3.1\times10^{3}$\\
BT code preparation & 147 & 13 & $0.90$ & $5.01\times10^{3}$ & 189 & 12 & $0.87$ & $5.7\times10^{3}$\\
Init. BB transversal meas. & 147 & 3 & $0.97$ & $5.59\times10^{3}$ & 189 & 3 & $\approx1$ & $6.3\times10^{3}$\\
BT transversal meas. & 189 & 8 & $0.56$ & $1.60\times10^{4}$ & 243 & 8 & $0.46$ & $2.3\times10^{4}$\\
First SE \& BB $Z$-error rejection & 252 & 12 & $0.95$ & $2.00\times10^{4}$ & 324 & 12 & $0.98$ & $2.8\times10^{4}$\\
Remaining SE \& postselection & 252 & 24 & $0.88$ & $2.95\times10^{4}$ & 324 & 24 & $0.82$ & $4.4\times10^{4}$\\
\end{tabular}
\end{ruledtabular}
\medskip
\begin{ruledtabular}
\begin{tabular}{l|rrrr|rrrr}
& \multicolumn{4}{c|}{$(n_{\mathrm{3D}},d_f)=(36,4)$} & \multicolumn{4}{c}{$(81,5)$}\\
Rejection stage & $Q$ & $D$ & $f$ & $2\overline V_{\le i}$ & $Q$ & $D$ & $f$ & $2\overline V_{\le i}$\\
\hline
Init. BB preparation & 216 & 24 & $0.68$ & $7.68\times10^{3}$ & 486 & 24 & $0.43$ & $2.7\times10^{4}$\\
BT code preparation & 252 & 13 & $0.81$ & $1.14\times10^{4}$ & 567 & 12 & $0.62$ & $3.8\times10^{4}$\\
Init. BB transversal meas. & 252 & 3 & $0.84$ & $1.48\times10^{4}$ & 567 & 3 & $0.97$ & $4.1\times10^{4}$\\
BT transversal meas. & 324 & 8 & $0.41$ & $5.08\times10^{4}$ & 729 & 8 & $0.16$ & $3.5\times10^{5}$\\
First SE \& BB $Z$-error rejection & 432 & 12 & $0.95$ & $5.88\times10^{4}$ & 972 & 12 & $0.77$ & $4.7\times10^{5}$\\
Remaining SE \& postselection & 432 & 24 & $0.79$ & $8.79\times10^{4}$ & 972 & 24 & $0.28$ & $1.7\times10^{6}$\\
\end{tabular}
\end{ruledtabular}
\caption{\label{Table:CCZresource}\textbf{Resource breakdown at each rejection stage.} The estimates use $p=0.1\%$. $Q$ and $D$ are the peak active qubits and first-try depth of each stage. $f$ is conditional acceptance rate: the first three rows of each panel show per-block acceptance rate, and subsequent rows apply to the full factory. $2\overline V_{\le i}$ is the cumulative expected active-qubit volume needed to pass the row, summed over all three blocks, before division by the two CCZ outputs.}
\end{table*}

In our CCZ state factory, we apply this accounting separately to the initialization of each code block, since these preparations can be repeated independently. We sum the expected preparation costs of the three blocks and include them in the initial segment of the joint CCZ and escape circuit. The estimate excludes routing and classical decoding latency. The survival fractions for this joint circuit are conditioned on three prepared input blocks. Writing the resulting expected volume per accepted pair as $\overline V_{\mathrm{factory}}$, the expected volume per CCZ is
\begin{align}
 \overline V_{\CCZ}^{\mathrm{direct}}
 =\frac{\overline V_{\mathrm{factory}}}{2}.
 \label{eq:volume-per-ccz}
\end{align}
Table~\ref{Table:CCZresource} groups operations by rejection stage and reports conditional acceptance probabilities $f$. Its $Q$ and $D$ columns give peak active-qubit counts and gate depths.

\subsection{\texorpdfstring{$T$}{T}-state based factory}\label{App:TcultivationCCZ}
\begin{figure}[!t]
    \centering
    \includegraphics[width=1.0\linewidth]{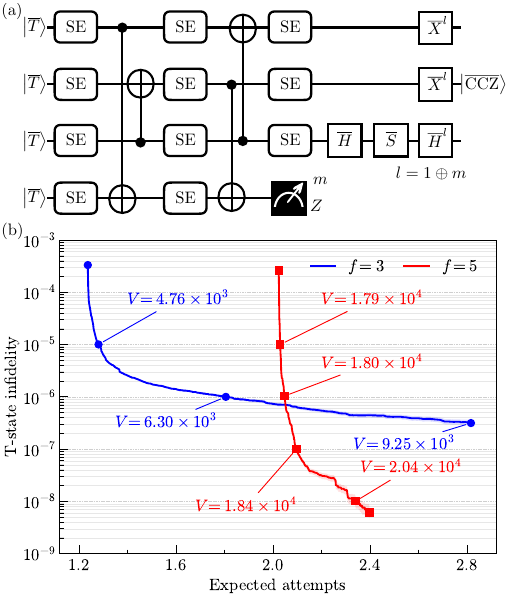}
    \caption{\textbf{$T$-state-based CCZ factory. }\textbf{(a)} $4T$ to CCZ conversion circuits. Four stored $T$ states can be converted into one CCZ state using Clifford circuits, mid-circuit measurements and adaptive feedback. One round of SE is interleaved with the transversal CNOTs to provide syndrome data for correlated decoding. \textbf{(b)} $T$-state cultivation infidelity and acceptance rate. The horizontal axis shows the expected number of attempts (i.e., the inverse of the acceptance rate), and the vertical axis shows the mean $T$-state infidelity. Expected spacetime volumes $V$ per accepted $T$ state are indicated at selected points.}
    \label{fig:TtoCCZ}
\end{figure}

We prepare four $T$ states using surface-code cultivation~\cite{sahay2025fold} and convert them into one CCZ state using the logical Clifford circuit shown in \figref{fig:TtoCCZ} (a). We perform the logical operations on the escape surface code, and the output CCZ state is stored in the three blocks of the surface code. We interleave SE with the transversal CNOTs to provide syndrome data for correlated decoding~\cite{PhysRevLett.133.240602, Zhou2025}. The spacetime volume of the conversion circuit is $\overline V_{\mathrm{conv}}=3.80\times10^4$ at distance $13$ and $5.06\times10^4$ at distance $15$, with a mean conversion depth of $36.5$ in both cases. The expected spacetime volume per CCZ state is
\begin{align}
 \overline V_{4T\to\CCZ}=4\overline V_T+\overline V_{\mathrm{conv}}.
\end{align}
Here, $\overline V_T$ is the expected spacetime volume per accepted $T$ state.

We simulate surface-code $T$-state cultivation to estimate the acceptance rate, mean output infidelity, and expected spacetime volume $\overline V_T$. We follow the $d_f=3$ and $d_f=5$ factory pipelines of Ref.~\cite{sahay2025fold} with two modifications. First, we use Caliper~\cite{mehta2026practicalcostmagicstate} to compute a complementary gap from the noisy syndrome history. Second, the non-Clifford operation is simulated exactly~\cite{fby6-xjbm, kim2026xtim}. We use the same noise model as for our CCZ factory, with noise strength $p=0.1\%$. Figure~\ref{fig:TtoCCZ}(b) shows the mean $T$-state infidelity versus the expected number of attempts (the inverse of the acceptance rate) as we vary the complementary-gap threshold. Expected spacetime volumes are shown at selected output infidelities.

\section{Details of codes}\label{App:CodeDetails}
\subsection{BT, base BB, and lifted BB codes}\label{app:logical-representatives}

\begin{table*}[!t]
\centering
\renewcommand{\arraystretch}{1.25}
\setlength{\tabcolsep}{4pt}
\begin{ruledtabular}
\begin{tabular}{lclll}
Code parameters & Algebra & Role or logical image
& \multicolumn{2}{c}{Defining polynomials}\\
\hline
$\llbracket21,3,(6,3)\rrbracket$ & $R_{7,1}$ & BT code
& \multicolumn{2}{l}{$a=1+x,\quad b=1+x^2,\quad c=1+x^3$}\\
$\llbracket28,1,1;(8,3)\rrbracket$ & $R_{7,2}$ & $\alpha$ (init. 1)
& $\widetilde b=1+x^2y$
& $\widetilde c=1+x^3y+x^5y+x^5$\\
$\llbracket28,1,1;(6,6)\rrbracket$ & $R_{7,2}$ & $\beta$ (init. 2)
& $\widetilde a=1+x^2+x+x^2y$
& $\widetilde c=1+x^3y$\\
$\llbracket28,1,1;(6,6)\rrbracket$ & $R_{7,2}$ & $\gamma$ (init. 3)
& $\widetilde b=1+x^2y$
& $\widetilde c=1+xy+x+x^3$\\
$\llbracket42,2,(8,8)\rrbracket$ & $R_{7,3}$ & $\mathcal I$ (escape 1--3)
& $\widetilde a=1+x^2y^2+x+x^2y$
& $\widetilde b=1+x^2+x^6y+x^6y^2$\\
\hline
$\llbracket27,3,(6,3)\rrbracket$ & $R_{3,3}$ & BT code
& \multicolumn{2}{l}{$a=x^2y+x^2y^2,\quad b=1+xy^2,\quad c=x+x^2y$}\\
$\llbracket36,1,1;(6,3)\rrbracket$ & $R_{3,6}$ & $\alpha$ (init. 1)
& $\widetilde a=x^2y+x^2y^5$
& $\widetilde b=1+xy^5$\\
$\llbracket36,1,1;(6,3)\rrbracket$ & $R_{3,6}$ & $\beta$ (init. 2)
& $\widetilde b=1+xy^2$
& $\widetilde c=x+x^2y$\\
$\llbracket36,1,1;(6,3)\rrbracket$ & $R_{3,6}$ & $\gamma$ (init. 3)
& $\widetilde c=x+x^2y^4$
& $\widetilde a=x^2y+x^2y^2$\\
$\llbracket54,2,(9,9)\rrbracket$ & $R_{3,9}$
& $\langle\beta,\gamma\rangle$ (escape 1)
& $\widetilde b=1+xy^8+x^2y+x^2y^4$
& $\widetilde c=y^2+y^5+xy^3+x^2y$\\
$\llbracket54,2,(9,9)\rrbracket$ & $R_{3,9}$
& $\langle\gamma,\alpha\rangle$ (escape 2)
& $\widetilde c=y+x+x^2y^2+x^2y^5$
& $\widetilde a=y^2+y^5+xy^3+x^2y$\\
$\llbracket54,2,(9,9)\rrbracket$ & $R_{3,9}$
& $\langle\alpha,\beta\rangle$ (escape 3)
& $\widetilde a=y+xy^3+xy^6+x^2y^2$
& $\widetilde b=y^3+xy+x^2y^2+x^2y^5$\\
\hline
$\llbracket36,3,(8,4)\rrbracket$ & $R_{3,4}$ & BT code
& \multicolumn{2}{l}{$a=1+xy,\quad b=1+y^3,\quad c=1+x^2y$}\\
$\llbracket48,1,1;(8,10)\rrbracket$ & $R_{3,8}$ & $\alpha$ (init. 1)
& $\widetilde a=1+xy+xy^4+x$
& $\widetilde c=1+y^3+x^2y^5+y^7$\\
$\llbracket48,1,1;(8,10)\rrbracket$ & $R_{3,8}$ & $\beta$ (init. 2)
& $\widetilde b=1+y^3+xy^4+x$
& $\widetilde c=1+x^2y^2+x^2y^5+x^2y^6$\\
$\llbracket48,1,1;(8,10)\rrbracket$ & $R_{3,8}$ & $\gamma$ (init. 3)
& $\widetilde a=1+xy+x^2y^2+x^2y^6$
& $\widetilde c=1+x^2+x^2y+x^2y^4$\\
$\llbracket72,2,(12,12)\rrbracket$ & $R_{9,4}$ & $\mathcal I$ (escape 1--3)
& $\widetilde a=1+xy+x^3y^3+x^6y^3$
& $\widetilde b=1+x^7y^3+xy^3+y^3$\\
\hline
$\llbracket81,3,(15,5)\rrbracket$ & $R_{3,9}$ & BT code
& \multicolumn{2}{l}{$a=1+xy^7,\quad b=1+xy^8,\quad c=1+y^2$}\\
$\llbracket108,1,1;(9,9)\rrbracket$ & $R_{3,18}$ & $\alpha$ (init. 1)
& $\widetilde a=1+xy^{16}$
& $\widetilde b=1+xy^{17}$\\
$\llbracket108,1,1;(9,9)\rrbracket$ & $R_{3,18}$ & $\beta$ (init. 2)
& $\widetilde b=1+xy^8$
& $\widetilde c=1+y^{11}$\\
$\llbracket108,1,1;(9,9)\rrbracket$ & $R_{3,18}$ & $\gamma$ (init. 3)
& $\widetilde c=1+y^2$
& $\widetilde a=1+xy^7$\\
$\llbracket162,2,(18,18)\rrbracket$ & $R_{3,27}$
& $\langle\beta,\gamma\rangle$ (escape 1)
& $\widetilde b=1+xy^{15}+xy^{24}+x^2y^2$
& $\widetilde c=1+x+xy^9+xy^{26}$\\
$\llbracket162,2,(18,18)\rrbracket$ & $R_{3,27}$
& $\langle\gamma,\alpha\rangle$ (escape 2)
& $\widetilde c=1+xy^{15}+xy^{24}+x^2y^2$
& $\widetilde a=1+y+y^{10}+x^2y$\\
$\llbracket162,2,(18,18)\rrbracket$ & $R_{3,27}$
& $\langle\alpha,\beta\rangle$ (escape 3)
& $\widetilde a=1+xy^{25}+x^2y^{13}+x^2y^{22}$
& $\widetilde b=1+x+xy^9+xy^{26}$\\
\end{tabular}
\end{ruledtabular}
\caption{
\label{tab:CodeDetails}
\textbf{The boundary maps of the BT/BB codes.}
We use $R_{l,m}=\F_2[x,y]/(x^l-1,y^m-1)$, and the variables $x,y$ generate $\Z_l,\Z_m$, respectively. The index-2 lifted BB codes are subsystem codes with one protected and one gauge qubit, written as $\llbracket n,k,r;(d_X,d_Z)\rrbracket$. The initialization rows identify the logical-$Z$ operators addressed by the protected BB logical qubits. The numbers following ``init.'' and ``escape'' label the BT block. Initialization checks have weight six for $n_{\mathrm{3D}}=21$, eight for $n_{\mathrm{3D}}=36$, and four for the other factories. Escape checks have weight eight. The polynomial labels identify the coupled BT factors; expressions use the coordinates of each BB component. The escape image $\mathcal I$ for $n_{\mathrm{3D}}=21,36$ is defined in \tabref{tab:escape-homomorphisms}. The BT codes with $n_{\mathrm{3D}} = 27, 81$ are from Ref.~\cite{li2026transversaldimensionjumpproduct}, and $n_{\mathrm{3D}} = 21, 36$ are from Ref.~\cite{jacob2026singleshotdecodingfaulttolerantgates}.
}
\end{table*}
Table~\ref{tab:CodeDetails} provides the boundary maps of the BT/BB codes used in our factory. All four factories use the same chain-complex construction over a group algebra $R_{l, m} = \F_2[\Z_l\times \Z_m] = \F_2[x,y]/(x^l-1,y^m-1)$ in Eqs.~\eqref{Eq:2Dchain} and~\eqref{eq:3D_complex}. The listed polynomials $a,b,c$ define the boundary maps of the underlying length-one chain complexes in Eq.~\eqref{Eq:BaseClassicalChain}. The base BB complexes use the pairs $(a,b)$, $(b,c)$, or $(c,a)$, and the two lifted polynomials in each BB row define the corresponding lifted codes.

Let $j=1,2,3$ denote the summands $Q_{100},Q_{010},Q_{001}$ of $Q_1^{\mathrm{ABC}}$, respectively. For $n_{\mathrm{3D}}=21,27,81$, choose reference operators $\overline X^{(j)}$ and $\overline Z^{(j)}$ with the common support polynomial $\sum_{a=0}^{l-1}\sum_{b=0}^{m-1}x^ay^b$ on summand $j$. These form conjugate logical pairs because $lm$ is odd. For $n_{\mathrm{3D}}=27,81$, the canonical basis is given by
\begin{align}
(\overline Z_\alpha,\overline Z_\beta,\overline Z_\gamma)
  &= (\overline Z^{(1)},\overline Z^{(2)}, \overline Z^{(3)}),\\
(\overline X_\alpha,\overline X_\beta,\overline X_\gamma)
  &= (\overline X^{(1)},\overline X^{(2)}, \overline{X}^{(3)}).
\end{align}
For $n_{\mathrm{3D}}=21$, the measured-parity basis is
\begin{align}
(\overline Z_\alpha,\overline Z_\beta,\overline Z_\gamma)
 &= (\overline Z^{(3)},\overline Z^{(1)}\overline Z^{(3)},\overline Z^{(2)}\overline Z^{(3)}),\\
(\overline X_\alpha,\overline X_\beta,\overline X_\gamma)
 &= (\overline X^{(1)}\overline X^{(2)}\overline X^{(3)},\overline X^{(1)},\overline X^{(2)}).
\end{align}

For $n_{\mathrm{3D}}=36$, let $X_j(f)$ and $Z_j(f)$ denote Pauli operators with support polynomial $f\in R_{3,4}$ on summand $j$, and write $s=1+x+x^2$. Explicit representatives of the measured-parity basis are
\begin{align}
 \overline Z_\alpha &= Z_1(1+x+xy+x^2y^2+y^3)Z_3(1),\\
 \overline Z_\beta &= Z_2(ys)Z_3(s),\\
 \overline Z_\gamma &= Z_1(s)Z_3(s).
\end{align}

The lifted-to-base chain maps are induced by group homomorphisms. For initialization, the defining groups in \tabref{tab:CodeDetails} admit the canonical quotient
\begin{align}
 \pi:\Z_{\widetilde l}\times\Z_{\widetilde m}&\longrightarrow\Z_l\times\Z_m,
 \nonumber\\
 \pi(u,v)&=(u\bmod l,v\bmod m).
\end{align}
Its linear extension defines $P:R_{\widetilde l,\widetilde m}\to R_{l,m}$. Applying $P$ componentwise gives the degreewise chain maps $(p_2,p_1,p_0)=(P,P\oplus P,P)$ from the initialization BB complex to the base BB complex.

For escape, each block uses the quotient homomorphism specified in \tabref{tab:escape-homomorphisms},
\begin{align}\label{eq:escape-group-homomorphism}
 \pi_b:\Z_{\widetilde l}\times\Z_{\widetilde m}&\longrightarrow\Z_l\times\Z_m.
\end{align}
Its linear extension defines $\mathcal P_b:R_{\widetilde l,\widetilde m}\to R_{l,m}$, giving $(p_2,p_1,p_0)=(\mathcal P_b,\mathcal P_b^{\oplus2},\mathcal P_b)$. In both cases, the projections send the lifted boundary polynomials to their corresponding base polynomials, ensuring the chain-map identities. Composing with the inclusion into the BT complex gives the lifted-BB-to-BT coupling in Eq.~\eqref{Eq:LiftedBBBTcoupling}.

\subsection{Null-homotopic map}
We specify the homotopies $h_1$ that generate the null-homotopic deformations described in \appref{App:chainmapDeformation}. Each $h_1$ acts on one qubit summand of the lifted BB complex and maps into one degree-two check summand of the BT complex; all other $h_i$ vanish. Table~\ref{tab:escape-homomorphisms} specifies the support, target summand, and monomial multiplier for each escape map.

Each fibre of $\pi_b$ contains three lifted preimages of a base group element. Within the selected source summand, the support of $h_1$ contains one representative from each fibre. For $n_{\mathrm{3D}}=27,81$, choose $\ell_1=\{x^u y^v:0\leq u<3,\ 0\leq v<n_{\mathrm{3D}}/9\}$. The translates $\ell_j=y^{(j-1)n_{\mathrm{3D}}/9}\ell_1$, $j=1,2,3$, partition the lifted group into three layers, each containing one representative from every fibre. For $n_{\mathrm{3D}}=21,36$, select $\ell_1=\{x^k y^k:0\leq k<n_{\mathrm{3D}}/3\}$ in all three blocks. The table selects a translated representative set $y^r\ell_1$ in the indicated source summand. This choice specifies the representatives labelled by the selected fibre coordinate in \figref{Fig:CMdeformation}.

Define $p_b$ by applying $\pi_b$ to the basis monomials in this support, setting it to zero on all other basis elements, and extending linearly. Multiplying its output by $t_1^{-1}$ and placing the result in the listed BT target summand defines $h_1$. Here $t=a,b,c$ for the BB components $\mathrm{BC},\mathrm{CA},\mathrm{AB}$, respectively, and $t_1$ is the selected monomial term of the omitted boundary polynomial $t=t_1+t_2$. Thus $t_1^{-1}$ generalizes the multiplier $c_1^{-1}$ in \figref{Fig:CMdeformation}.

\begin{table}[!t]
\centering
\renewcommand{\arraystretch}{1.25}
\setlength{\tabcolsep}{2pt}
\begin{ruledtabular}
\begin{tabular}{cccccccc}
$n_{\mathrm{3D}}$ & \shortstack{Logical\\image} & $\pi_b(x)$ & $\pi_b(y)$ & \multicolumn{2}{c}{$\operatorname{supp}p_b$} & \shortstack{$h_1$\\target} & $t_1^{-1}$\\
\hline
$21$ & $\mathcal I$ & $x$ & $1$ & $\widetilde Q_{10}^{\mathrm{AB}}$ & $\ell_1$ & $Q_{101}$ & $1$\\
    & $\mathcal I$ & $x$ & $1$ & $\widetilde Q_{01}^{\mathrm{AB}}$ & $\ell_1$ & $Q_{011}$ & $1$\\
    & $\mathcal I$ & $x$ & $1$ & $\widetilde Q_{10}^{\mathrm{AB}}$ & $\ell_1$ & $Q_{101}$ & $1$\\
\hline
$27$ & $\langle\beta,\gamma\rangle$ & $x$ & $y$ & $\widetilde Q_{10}^{\mathrm{BC}}$ & $y^{-1}\ell_1$ & $Q_{110}$ & $xy^2$\\
    & $\langle\gamma,\alpha\rangle$ & $x^2y$ & $x$ & $\widetilde Q_{10}^{\mathrm{CA}}$ & $y^{-2}\ell_1$ & $Q_{011}$ & $1$\\
    & $\langle\alpha,\beta\rangle$ & $x^2$ & $x^2y^2$ & $\widetilde Q_{01}^{\mathrm{AB}}$ & $y^{-3}\ell_1$ & $Q_{011}$ & $x^2$\\
\hline
$36$ & $\mathcal I$ & $x$ & $y$ & $\widetilde Q_{01}^{\mathrm{AB}}$ & $\ell_1$ & $Q_{011}$ & $1$\\
     & $\mathcal I$ & $x$ & $y$ & $\widetilde Q_{10}^{\mathrm{AB}}$ & $\ell_1$ & $Q_{101}$ & $1$\\
     & $\mathcal I$ & $x$ & $y$ & $\widetilde Q_{01}^{\mathrm{AB}}$ & $\ell_1$ & $Q_{011}$ & $1$\\
\hline
$81$ & $\langle\beta,\gamma\rangle$ & $xy^3$ & $xy$ & $\widetilde Q_{10}^{\mathrm{BC}}$ & $y^{-10}\ell_1$ & $Q_{110}$ & $1$\\
    & $\langle\gamma,\alpha\rangle$ & $xy^6$ & $x^2y^4$ & $\widetilde Q_{10}^{\mathrm{CA}}$ & $y^{-10}\ell_1$ & $Q_{011}$ & $1$\\
    & $\langle\alpha,\beta\rangle$ & $x$ & $y$ & $\widetilde Q_{10}^{\mathrm{AB}}$ & $y^{-8}\ell_1$ & $Q_{101}$ & $1$\\
\end{tabular}
\end{ruledtabular}
\caption{\label{tab:escape-homomorphisms}
\textbf{Escape code group homomorphisms and chain map deformations.}
For each $n_{\mathrm{3D}}$, the rows correspond to blocks $b=1,2,3$. The generator images specify the escape projections from $\Z_{\tilde l}\times\Z_{\tilde m}$ to $\Z_l\times\Z_m$. For the $21$- and $36$-qubit BT codes, the three escape codes are identical and have the logical image $\mathcal I=\langle\alpha+\beta,\alpha+\gamma\rangle$. The last four columns give the source summand and representative set supporting $p_b$, the target of $h_1$ in $Q_2$, and its multiplier.}
\end{table}

\section{Simulation}\label{app}
We describe the fast simulation of adaptive measurements in more detail. We consider measuring the Pauli $X$ operators $X_i$ on the noisy state $\ket{\psi} = U_{\mathrm{Cl}}\ket{\overline{\CCZ}}^{\otimes 2}$, where $U_\mathrm{Cl}$ represents the propagated stochastic Clifford errors. The measurement outcome $s_i\in\{+1,-1\}$ is sampled with probability
\begin{align}
    p(s_i\mid\psi)=\ev{(I + s_iX_i)/2}{\psi}, 
\end{align}
and the state collapses to 
\begin{align}
    \frac{1}{\sqrt{p(s_i\mid\psi)}}\frac{I + s_i X_i}2 \ket{\psi}.
\end{align}
In the transversal readout, we simultaneously measure a set of Pauli $X$ operators $\{X_i\}_i$, sample their outcomes according to the Born rule, and update the state accordingly.

For each run with a sampled Clifford error $U_\mathrm{Cl}$, this is equivalent to measuring the Pauli operators
\begin{align}
 P_i=U_{\mathrm{Cl}}^\dagger X_iU_{\mathrm{Cl}}
\end{align}
on the ideal output state
\begin{align}
    \ket{\psi}_{\mathrm{clean}} = \ket{\overline{\mathrm{CCZ}}}^{\otimes 2} =(\ket{\overline{+++}} - \ket{\overline{111}}/\sqrt{2})^{\otimes 2}.
\end{align}

The four stabilizer components of the ideal output share $n-6$ independent stabilizers, leaving a six-qubit nonstabilizer subsystem. Stabilizer-tableau elimination~\cite{aaronson2008improvedsimulationstabilizercircuits} separates the measured Pauli products into the shared stabilizers, the shared destabilizers, and the six-qubit nonstabilizer subsystem. A product in the shared stabilizer group has a fixed outcome. Measuring a product that anticommutes with a shared generator yields a random outcome and updates the shared stabilizers after projection. A product that commutes with every shared generator but is outside the stabilizer group acts on the remaining six-qubit subsystem. We evaluate these residual measurements using at most 64 complex amplitudes, retaining interference between the four components. For a commuting measurement layer, we reduce the measured operators to independent products and sample their conditional Born probabilities in sequence, updating the state after each outcome; this preserves correlations between the readout bits.

\clearpage
\bibliography{refs}

@misc{lee2026designinggroupvaluedcodesregular,
      title={Designing Group-Valued Codes with Full Regular Low-Weight Bases}, 
      author={Jong Yeon Lee},
      year={2026},
      eprint={2610.06820},
      archivePrefix={arXiv},
      primaryClass={quant-ph},
      url={https://arxiv.org/abs/2610.06820}, 
}

@article{Pordes2007OpenScienceGrid,
  author  = {Pordes, R. and Petravick, D. and Kramer, B. and Olson, D. and Livny, M. and Roy, A. and Avery, P. and Blackburn, K. and Wenaus, T. and W{\"u}rthwein, F. and Foster, I. and Gardner, R. and Wilde, M. and Blatecky, A. and McGee, J. and Quick, R.},
  title   = {The Open Science Grid},
  journal = {Journal of Physics: Conference Series},
  volume  = {78},
  pages   = {012057},
  year    = {2007},
  doi     = {10.1088/1742-6596/78/1/012057},
  url     = {https://doi.org/10.1088/1742-6596/78/1/012057}
}

@inproceedings{Sfiligoi2009GlideinWMS,
  author    = {Sfiligoi, I. and Bradley, D. C. and Holzman, B. and Mhashilkar, P. and Padhi, S. and Wurthwein, F.},
  title     = {The Pilot Way to Grid Resources Using {glideinWMS}},
  booktitle = {2009 WRI World Congress on Computer Science and Information Engineering},
  volume    = {2},
  pages     = {428--432},
  year      = {2009},
  doi       = {10.1109/CSIE.2009.950},
  url       = {https://doi.org/10.1109/CSIE.2009.950}
}

@misc{OSG2006OSPool,
  author       = {{OSG}},
  title        = {{OSPool}},
  year         = {2006},
  howpublished = {OSG},
  doi          = {10.21231/906P-4D78},
  url          = {https://doi.org/10.21231/906P-4D78}
}

@misc{OSG2015OSDF,
  author       = {{OSG}},
  title        = {{Open Science Data Federation}},
  year         = {2015},
  howpublished = {OSG},
  doi          = {10.21231/0KVZ-VE57},
  url          = {https://doi.org/10.21231/0KVZ-VE57}
}

@article{bravyi2005universal,
      title={Universal quantum computation with ideal Clifford gates and noisy ancillas},
      author={Bravyi, Sergey and Kitaev, Alexei},
      journal={Physical Review A},
      volume={71},
      pages={022316},
      year={2005},
      doi={10.1103/PhysRevA.71.022316}
}

@article{PhysRevA.86.052329,
  title = {Magic-state distillation with low overhead},
  author = {Bravyi, Sergey and Haah, Jeongwan},
  journal = {Phys. Rev. A},
  volume = {86},
  issue = {5},
  pages = {052329},
  numpages = {10},
  year = {2012},
  month = {Nov},
  publisher = {American Physical Society},
  doi = {10.1103/PhysRevA.86.052329},
  url = {https://link.aps.org/doi/10.1103/PhysRevA.86.052329}
}

@misc{xu2026distillingmagicstatesbicycle,
      title={Distilling Magic States in the Bicycle Architecture}, 
      author={Shifan Xu and Kun Liu and Patrick Rall and Zhiyang He and Yongshan Ding},
      year={2026},
      eprint={2602.20546},
      archivePrefix={arXiv},
      primaryClass={quant-ph},
      url={https://arxiv.org/abs/2602.20546}, 
}

@article{sv11-f78l,
  title = {Simple, Efficient, and Generic Postselection Decoding for qLDPC Codes},
  author = {Xie, Haipeng and Yoshioka, Nobuyuki and Tsubouchi, Kento and Li, Ying},
  journal = {Phys. Rev. Lett.},
  volume = {137},
  issue = {11},
  pages = {110601},
  numpages = {7},
  year = {2026},
  month = {Sep},
  publisher = {American Physical Society},
  doi = {10.1103/sv11-f78l},
  url = {https://link.aps.org/doi/10.1103/sv11-f78l}
}

@article{PhysRevX.15.021065,
  title = {Fast and Parallelizable Logical Computation with Homological Product Codes},
  author = {Xu, Qian and Zhou, Hengyun and Zheng, Guo and Bluvstein, Dolev and Ataides, J. Pablo Bonilla and Lukin, Mikhail D. and Jiang, Liang},
  journal = {Phys. Rev. X},
  volume = {15},
  issue = {2},
  pages = {021065},
  numpages = {34},
  year = {2025},
  month = {May},
  publisher = {American Physical Society},
  doi = {10.1103/PhysRevX.15.021065},
  url = {https://link.aps.org/doi/10.1103/PhysRevX.15.021065}
}

@misc{hirasaki2026liftingliftedproductcodes,
      title={Lifting Lifted Product Codes}, 
      author={Yuta Hirasaki and Jong Yeon Lee},
      year={2026},
      eprint={2607.28621},
      archivePrefix={arXiv},
      primaryClass={quant-ph},
      url={https://arxiv.org/abs/2607.28621}, 
}

@article{huang2023homomorphic,
      title={Homomorphic logical measurements},
      author={Shilin Huang and Tomas Jochym-O'Connor and Theodore J. Yoder},
      journal={PRX Quantum},
      volume={4},
      pages={030301},
      year={2023},
      doi={10.1103/PRXQuantum.4.030301}
}

@misc{gidney2024magic,
      title={Magic state cultivation: growing T states as cheap as CNOT gates},
      author={Craig Gidney and Noah Shutty and Cody Jones},
      year={2024},
      eprint={2409.17595},
      archivePrefix={arXiv},
      primaryClass={quant-ph}
}

@article{sahay2025fold,
  title = {Fold-transversal surface code cultivation},
  author = {Sahay, Kaavya and Tsai, Pei-Kai and Chang, Kathleen (Katie) and Su, Qile and Smith, Thomas B. and Singh, Shraddha and Puri, Shruti},
  journal = {PRX Quantum},
  volume = {7},
  issue = {3},
  pages = {033006},
  numpages = {26},
  year = {2026},
  month = {Jul},
  publisher = {American Physical Society},
  doi = {10.1103/gpvl-lg4c},
  url = {https://link.aps.org/doi/10.1103/gpvl-lg4c}
}

@Article{Chamberland2020,
  author    = {Chamberland, Christopher and Noh, Kyungjoo},
  journal   = {npj Quantum Information},
  title     = {Very low overhead fault-tolerant magic state preparation using redundant ancilla encoding and flag qubits},
  year      = {2020},
  issn      = {2056-6387},
  month     = Oct,
  number    = {1},
  volume    = {6},
  doi       = {10.1038/s41534-020-00319-5},
  publisher = {Springer Science and Business Media LLC},
}

@misc{muller2025relay,
      title={Improved belief propagation is sufficient for real-time decoding of quantum memory},
      author={Timo M{\"u}ller and Thomas Alexander and Michael E. Beverland and Michael B{\"u}hler and Benjamin R. Johnson and Thomas Maurer and Daniel Vandeth},
      year={2025},
      eprint={2506.01779},
      archivePrefix={arXiv},
      primaryClass={quant-ph}
}

@misc{gong2026magicstatedistillationcodes,
      title={Magic State Distillation via Codes over Binary Extension Fields}, 
      author={Anqi Gong and Christopher A. Pattison and Patrick Rall and Adam Wills},
      year={2026},
      eprint={2608.09727},
      archivePrefix={arXiv},
      primaryClass={quant-ph},
      url={https://arxiv.org/abs/2608.09727}, 
}

@misc{symons2026sequencesbivariatebicyclecodes,
      title={Sequences of Bivariate Bicycle Codes from Covering Graphs}, 
      author={Benjamin C. B. Symons and Abhishek Rajput and Dan E. Browne},
      year={2026},
      eprint={2511.13560},
      archivePrefix={arXiv},
      primaryClass={quant-ph},
      url={https://arxiv.org/abs/2511.13560}, 
}

@Article{Lee2026,
  author    = {Lee, Seok-Hyung and English, Lucas H. and Bartlett, Stephen D.},
  journal   = {npj Quantum Information},
  title     = {Efficient post-selection for general quantum LDPC Codes},
  year      = {2026},
  issn      = {2056-6387},
  month     = Apr,
  number    = {1},
  volume    = {12},
  doi       = {10.1038/s41534-026-01242-x},
  publisher = {Springer Science and Business Media LLC},
}

@Article{Gidney2025,
  author    = {Gidney, Craig and Newman, Michael and Brooks, Peter and Jones, Cody},
  journal   = {Nature Communications},
  title     = {Yoked surface codes},
  year      = {2025},
  issn      = {2041-1723},
  month     = May,
  number    = {1},
  volume    = {16},
  doi       = {10.1038/s41467-025-59714-1},
  publisher = {Springer Science and Business Media LLC},
}

@article{fby6-xjbm,
  title = {Efficient Simulation of Logical Magic State Preparation Protocols},
  author = {Surti, Samyak and Daguerre, Lucas and Kim, Isaac H.},
  journal = {PRX Quantum},
  volume = {7},
  issue = {2},
  pages = {020329},
  numpages = {32},
  year = {2026},
  month = {May},
  publisher = {American Physical Society},
  doi = {10.1103/fby6-xjbm},
  url = {https://link.aps.org/doi/10.1103/fby6-xjbm}
}

@article{PhysRevResearch.7.023080,
  title = {Code switching revisited: Low-overhead magic state preparation using color codes},
  author = {Daguerre, Lucas and Kim, Isaac H.},
  journal = {Phys. Rev. Res.},
  volume = {7},
  issue = {2},
  pages = {023080},
  numpages = {12},
  year = {2025},
  month = {Apr},
  publisher = {American Physical Society},
  doi = {10.1103/PhysRevResearch.7.023080},
  url = {https://link.aps.org/doi/10.1103/PhysRevResearch.7.023080}
}

@article{dck4-x9c2,
  title = {Experimental Demonstration of High-Fidelity Logical Magic States from Code Switching},
  author = {Daguerre, Lucas and Blume-Kohout, Robin and Brown, Natalie C. and Hayes, David and Kim, Isaac H.},
  journal = {Phys. Rev. X},
  volume = {15},
  issue = {4},
  pages = {041008},
  numpages = {11},
  year = {2025},
  month = {Oct},
  publisher = {American Physical Society},
  doi = {10.1103/dck4-x9c2},
  url = {https://link.aps.org/doi/10.1103/dck4-x9c2}
}

@misc{guemard2024liftsquantumcsscodes,
      title={Lifts of quantum CSS codes}, 
      author={Virgile Guemard},
      year={2024},
      eprint={2404.16736},
      archivePrefix={arXiv},
      primaryClass={quant-ph},
      url={https://arxiv.org/abs/2404.16736}, 
}

@misc{chen2025efficientmagicstatecultivation,
      title={Efficient Magic State Cultivation on $\mathbb{RP}^2$}, 
      author={Zi-Han Chen and Ming-Cheng Chen and Chao-Yang Lu and Jian-Wei Pan},
      year={2025},
      eprint={2503.18657},
      archivePrefix={arXiv},
      primaryClass={quant-ph},
      url={https://arxiv.org/abs/2503.18657}, 
}

@misc{claes2025cultivatingtstatessurface,
      title={Cultivating T states on the surface code with only two-qubit gates}, 
      author={Jahan Claes},
      year={2025},
      eprint={2509.05232},
      archivePrefix={arXiv},
      primaryClass={quant-ph},
      url={https://arxiv.org/abs/2509.05232}, 
}

@Article{Wills2025,
  author    = {Wills, Adam and Hsieh, Min-Hsiu and Yamasaki, Hayata},
  journal   = {Nature Physics},
  title     = {Constant-overhead magic state distillation},
  year      = {2025},
  issn      = {1745-2481},
  month     = sep,
  number    = {11},
  pages     = {1842--1846},
  volume    = {21},
  doi       = {10.1038/s41567-025-03026-0},
  publisher = {Springer Science and Business Media LLC},
}

@inproceedings{golowich2024asymptoticallygoodquantumcodes,
author = {Golowich, Louis and Guruswami, Venkatesan},
title = {Asymptotically Good Quantum Codes with Transversal Non-Clifford Gates},
year = {2025},
isbn = {9798400715105},
publisher = {Association for Computing Machinery},
address = {New York, NY, USA},
url = {https://doi.org/10.1145/3717823.3718234},
doi = {10.1145/3717823.3718234},
booktitle = {Proceedings of the 57th Annual ACM Symposium on Theory of Computing},
pages = {707–717},
numpages = {11},
location = {Prague, Czechia},
series = {STOC '25}
}

@inproceedings{10.1145/3717823.3718186,
author = {Nguyen, Quynh T.},
title = {Good Binary Quantum Codes with Transversal CCZ Gate},
year = {2025},
isbn = {9798400715105},
publisher = {Association for Computing Machinery},
address = {New York, NY, USA},
url = {https://doi.org/10.1145/3717823.3718186},
doi = {10.1145/3717823.3718186},
booktitle = {Proceedings of the 57th Annual ACM Symposium on Theory of Computing},
pages = {697–706},
numpages = {10},
location = {Prague, Czechia},
series = {STOC '25}
}

@article{2481602.2481604,
author = {Meier, Adam M. and Eastin, Bryan and Knill, Emanuel},
title = {Magic-state distillation with the four-qubit code},
year = {2013},
issue_date = {March 2013},
publisher = {Rinton Press, Incorporated},
address = {Paramus, NJ},
volume = {13},
number = {3–4},
issn = {1533-7146},
journal = {Quantum Info. Comput.},
month = mar,
pages = {195–209},
numpages = {15}
}

@inproceedings{10.1145/3717823.3718318,
author = {Nguyen, Quynh T. and Pattison, Christopher A.},
title = {Quantum Fault Tolerance with Constant-Space and Logarithmic-Time Overheads},
year = {2025},
isbn = {9798400715105},
publisher = {Association for Computing Machinery},
address = {New York, NY, USA},
url = {https://doi.org/10.1145/3717823.3718318},
doi = {10.1145/3717823.3718318},
booktitle = {Proceedings of the 57th Annual ACM Symposium on Theory of Computing},
pages = {730–737},
numpages = {8},
location = {Prague, Czechia},
series = {STOC '25}
}

@article{PhysRevA.87.042305,
  title = {Multilevel distillation of magic states for quantum computing},
  author = {Jones, Cody},
  journal = {Phys. Rev. A},
  volume = {87},
  issue = {4},
  pages = {042305},
  numpages = {8},
  year = {2013},
  month = {Apr},
  publisher = {American Physical Society},
  doi = {10.1103/PhysRevA.87.042305},
  url = {https://link.aps.org/doi/10.1103/PhysRevA.87.042305}
}

@misc{itogawa2026zerolevelcczdistillation,
      title={Zero-level $CCZ$ Distillation}, 
      author={Tomohiro Itogawa and Yutaka Hirano and Yutaro Akahoshi and Keisuke Fujii},
      year={2026},
      eprint={2605.21867},
      archivePrefix={arXiv},
      primaryClass={quant-ph},
      url={https://arxiv.org/abs/2605.21867}, 
}

@article{thxx-njr6,
  title = {Efficient Magic State Distillation by Zero-Level Distillation},
  author = {Itogawa, Tomohiro and Takada, Yugo and Hirano, Yutaka and Fujii, Keisuke},
  journal = {PRX Quantum},
  volume = {6},
  issue = {2},
  pages = {020356},
  numpages = {12},
  year = {2025},
  month = {Jun},
  publisher = {American Physical Society},
  doi = {10.1103/thxx-njr6},
  url = {https://link.aps.org/doi/10.1103/thxx-njr6}
}

@article{Haah2017magicstate,
  doi = {10.22331/q-2017-10-03-31},
  url = {https://doi.org/10.22331/q-2017-10-03-31},
  title = {Magic state distillation with low space overhead and optimal asymptotic input count},
  author = {Haah, Jeongwan and Hastings, Matthew B. and Poulin, D. and Wecker, D.},
  journal = {{Quantum}},
  issn = {2521-327X},
  publisher = {{Verein zur F{\"{o}}rderung des Open Access Publizierens in den Quantenwissenschaften}},
  volume = {1},
  pages = {31},
  month = oct,
  year = {2017}
}

@article{PhysRevLett.120.050504,
  title = {Distillation with Sublogarithmic Overhead},
  author = {Hastings, Matthew B. and Haah, Jeongwan},
  journal = {Phys. Rev. Lett.},
  volume = {120},
  issue = {5},
  pages = {050504},
  numpages = {3},
  year = {2018},
  month = {Jan},
  publisher = {American Physical Society},
  doi = {10.1103/PhysRevLett.120.050504},
  url = {https://link.aps.org/doi/10.1103/PhysRevLett.120.050504}
}

@article{PhysRevLett.123.070507,
  title = {Towards Low Overhead Magic State Distillation},
  author = {Krishna, Anirudh and Tillich, Jean-Pierre},
  journal = {Phys. Rev. Lett.},
  volume = {123},
  issue = {7},
  pages = {070507},
  numpages = {4},
  year = {2019},
  month = {Aug},
  publisher = {American Physical Society},
  doi = {10.1103/PhysRevLett.123.070507},
  url = {https://link.aps.org/doi/10.1103/PhysRevLett.123.070507}
}

@misc{cervia2025magicstatedistillationusing,
      title={Magic State Distillation using Asymptotically Good Codes on Qudits}, 
      author={Michael J. Cervia and Henry Lamm and Diyi Liu and Edison M. Murairi and Shuchen Zhu},
      year={2025},
      eprint={2512.21874},
      archivePrefix={arXiv},
      primaryClass={quant-ph},
      url={https://arxiv.org/abs/2512.21874}, 
}

@article{Litinski2019magicstate,
  doi = {10.22331/q-2019-12-02-205},
  url = {https://doi.org/10.22331/q-2019-12-02-205},
  title = {Magic {S}tate {D}istillation: {N}ot as {C}ostly as {Y}ou {T}hink},
  author = {Litinski, Daniel},
  journal = {{Quantum}},
  issn = {2521-327X},
  publisher = {{Verein zur F{\"{o}}rderung des Open Access Publizierens in den Quantenwissenschaften}},
  volume = {3},
  pages = {205},
  month = dec,
  year = {2019}
}

@article{PhysRevLett.133.240602,
  title = {Correlated Decoding of Logical Algorithms with Transversal Gates},
  author = {Cain, Madelyn and Zhao, Chen and Zhou, Hengyun and Meister, Nadine and Ataides, J. Pablo Bonilla and Jaffe, Arthur and Bluvstein, Dolev and Lukin, Mikhail D.},
  journal = {Phys. Rev. Lett.},
  volume = {133},
  issue = {24},
  pages = {240602},
  numpages = {7},
  year = {2024},
  month = {Dec},
  publisher = {American Physical Society},
  doi = {10.1103/PhysRevLett.133.240602},
  url = {https://link.aps.org/doi/10.1103/PhysRevLett.133.240602}
}

@Article{Zhou2025,
  author    = {Zhou, Hengyun and Zhao, Chen and Cain, Madelyn and Bluvstein, Dolev and Maskara, Nishad and Duckering, Casey and Hu, Hong-Ye and Wang, Sheng-Tao and Kubica, Aleksander and Lukin, Mikhail D.},
  journal   = {Nature},
  title     = {Low-overhead transversal fault tolerance for universal quantum computation},
  year      = {2025},
  issn      = {1476-4687},
  month     = sep,
  number    = {8084},
  pages     = {303--308},
  volume    = {646},
  doi       = {10.1038/s41586-025-09543-5},
  publisher = {Springer Science and Business Media LLC},
}

@misc{sahay2026computingqldpccodesclimbing,
      title={Computing with qLDPC Codes by Climbing the Chain Map Hierarchy}, 
      author={Rahul Sahay and David M. Long and Vedika Khemani},
      year={2026},
      eprint={2609.02999},
      archivePrefix={arXiv},
      primaryClass={quant-ph},
      url={https://arxiv.org/abs/2609.02999}, 
}

@misc{berent2026liftedsurgeryfastprocessing,
      title={Lifted surgery: Fast processing with QLDPC codes}, 
      author={Lucas Berent and Lawrence Z. Cohen and Armanda O. Quintavalle},
      year={2026},
      eprint={2609.11723},
      archivePrefix={arXiv},
      primaryClass={quant-ph},
      url={https://arxiv.org/abs/2609.11723}, 
}

@book{weibel1994introduction,
  title={An Introduction to Homological Algebra},
  author={Weibel, C.A.},
  isbn={9780521559874},
  lccn={93015649},
  series={Cambridge Studies in Advanced Mathematics},
  url={https://books.google.ne/books?id=flm-dBXfZ_gC},
  year={1994},
  publisher={Cambridge University Press}
}

@book{hatcher2002algebraic,
  title={Algebraic topology},
  author={Hatcher, Allen},
  year={2002},
  publisher={Cambridge University Press}
}

@Article{Maan2026,
  author    = {Maan, Arshpreet Singh and Garcia Herrero, Francisco Miguel and Paler, Alexandru and Savin, Valentin},
  journal   = {Nature Communications},
  title     = {Decoding correlated errors in quantum LDPC codes},
  year      = {2026},
  issn      = {2041-1723},
  month     = Mar,
  number    = {1},
  volume    = {17},
  doi       = {10.1038/s41467-026-70556-3},
  publisher = {Springer Science and Business Media LLC},
}

@misc{beni2025tesseractsearchbaseddecoderquantum,
      title={Tesseract: A Search-Based Decoder for Quantum Error Correction}, 
      author={Laleh Aghababaie Beni and Oscar Higgott and Noah Shutty},
      year={2025},
      eprint={2503.10988},
      archivePrefix={arXiv},
      primaryClass={quant-ph},
      url={https://arxiv.org/abs/2503.10988}, 
}

@misc{bhardwaj2026highrateqldpcprocessors,
      title={High-rate qLDPC processors}, 
      author={Aditya Bhardwaj and Muzhou Ma and Nadine Meister and Robbie King and Dolev Bluvstein and John Preskill and Madelyn Cain and Qian Xu and Hsin-Yuan Huang},
      year={2026},
      eprint={2607.28795},
      archivePrefix={arXiv},
      primaryClass={quant-ph},
      url={https://arxiv.org/abs/2607.28795}, 
}

@misc{wang2026checkweightconstrainedquantumcodesbounds,
      title={Check-weight-constrained quantum codes: Bounds and examples}, 
      author={Lily Wang and Andy Zeyi Liu and Ray Li and Aleksander Kubica and Shouzhen Gu},
      year={2026},
      eprint={2601.15446},
      archivePrefix={arXiv},
      primaryClass={quant-ph},
      url={https://arxiv.org/abs/2601.15446}, 
}

@misc{lee2026designprinciplesultrahighratequantum,
      title={Design Principles for Ultra-High-Rate Quantum Codes}, 
      author={Jong Yeon Lee and Koki Okada and Nishad Maskara and Kenta Kasai and Hengyun Zhou},
      year={2026},
      eprint={2609.30069},
      archivePrefix={arXiv},
      primaryClass={quant-ph},
      url={https://arxiv.org/abs/2609.30069}, 
}

@article{vrty-qs5h,
  title = {Universal Fault-Tolerant Quantum Computation in 2D without Getting Tied in Knots},
  author = {Davydova, Margarita and Bauer, Andreas and Magdalena de la Fuente, Julio C. and Webster, Mark and Williamson, Dominic J. and Brown, Benjamin J.},
  journal = {Phys. Rev. X},
  volume = {16},
  issue = {3},
  pages = {031001},
  numpages = {62},
  year = {2026},
  month = {Jul},
  publisher = {American Physical Society},
  doi = {10.1103/vrty-qs5h},
  url = {https://link.aps.org/doi/10.1103/vrty-qs5h}
}

@article{f7fz-6zsy,
  title = {Constant-Depth Magic State Cultivation with Clifford Measurements by Gauging},
  author = {Het{\'e}nyi, Bence and Brown, Benjamin J. and Williamson, Dominic J.},
  journal = {Phys. Rev. Lett.},
  volume = {137},
  issue = {14},
  pages = {140601},
  numpages = {6},
  year = {2026},
  month = {Sep},
  publisher = {American Physical Society},
  doi = {10.1103/f7fz-6zsy},
  url = {https://link.aps.org/doi/10.1103/f7fz-6zsy}
}

@misc{hirano2026efficientmagicstatecultivation,
      title={Efficient magic state cultivation with lattice surgery}, 
      author={Yutaka Hirano and Riki Toshio and Tomohiro Itogawa and Keisuke Fujii},
      year={2026},
      eprint={2510.24615},
      archivePrefix={arXiv},
      primaryClass={quant-ph},
      url={https://arxiv.org/abs/2510.24615}, 
}

@misc{mehta2026practicalcostmagicstate,
      title={The practical cost of magic state cultivation}, 
      author={Rohan Mehta and Varun Menon and Hengyun Zhou and Mikhail D. Lukin and J. Pablo Bonilla Ataides},
      year={2026},
      eprint={2610.06828},
      archivePrefix={arXiv},
      primaryClass={quant-ph},
      url={https://arxiv.org/abs/2610.06828}, 
}

@article{bauer2026planarfaulttolerantcircuitsnoncliffo,
  title = {Planar Fault-Tolerant Circuits for Non-Clifford Gates on the 2D Color Code},
  author = {Bauer, Andreas and Magdalena de la Fuente, Julio C.},
  journal = {PRX Quantum},
  volume = {7},
  issue = {2},
  pages = {020367},
  numpages = {57},
  year = {2026},
  month = {Jun},
  publisher = {American Physical Society},
  doi = {10.1103/p7c9-x1m9},
  url = {https://link.aps.org/doi/10.1103/p7c9-x1m9}
}

@misc{gidney2025factor2048bitrsa,
      title={How to factor 2048 bit RSA integers with less than a million noisy qubits},
      author={Craig Gidney},
      year={2025},
      eprint={2505.15917},
      archivePrefix={arXiv},
      primaryClass={quant-ph},
      url={https://arxiv.org/abs/2505.15917},
}

@misc{guémard2025moderatelengthliftedquantumtanner,
      title={Moderate-length lifted quantum Tanner codes}, 
      author={Virgile Guémard and Gilles Zémor},
      year={2025},
      eprint={2502.20297},
      archivePrefix={arXiv},
      primaryClass={quant-ph},
      url={https://arxiv.org/abs/2502.20297}, 
}

@article{gidney2018halvingcostquantumaddition,
  doi = {10.22331/q-2018-06-18-74},
  url = {https://doi.org/10.22331/q-2018-06-18-74},
  title = {Halving the cost of quantum addition},
  author = {Gidney, Craig},
  journal = {{Quantum}},
  issn = {2521-327X},
  publisher = {{Verein zur F{\"{o}}rderung des Open Access Publizierens in den Quantenwissenschaften}},
  volume = {2},
  pages = {74},
  month = jun,
  year = {2018}
}

@article{eastin2009restrictionstransversalencodedquantum,
  title = {Restrictions on Transversal Encoded Quantum Gate Sets},
  author = {Eastin, Bryan and Knill, Emanuel},
  journal = {Phys. Rev. Lett.},
  volume = {102},
  issue = {11},
  pages = {110502},
  numpages = {4},
  year = {2009},
  month = {Mar},
  publisher = {American Physical Society},
  doi = {10.1103/PhysRevLett.102.110502},
  url = {https://link.aps.org/doi/10.1103/PhysRevLett.102.110502}
}

@misc{min2026directcultivationentangledcsrangle,
      title={Direct Cultivation of Entangled $|CS\rangle$ Magic States}, 
      author={Gunsik Min and Jun Heo},
      year={2026},
      eprint={2609.09993},
      archivePrefix={arXiv},
      primaryClass={quant-ph},
      url={https://arxiv.org/abs/2609.09993}, 
}

@misc{min2026codedcliffordmeasurementsmultiqubit,
      title={Coded Clifford Measurements for Multiqubit Magic-State Cultivation},
      author={Gunsik Min and Jun Heo},
      year={2026},
      eprint={2609.09994},
      archivePrefix={arXiv},
      primaryClass={quant-ph},
      url={https://arxiv.org/abs/2609.09994},
}

@article{Rossi_2013,
doi = {10.1088/1367-2630/15/11/113022},
url = {https://doi.org/10.1088/1367-2630/15/11/113022},
year = {2013},
month = {nov},
publisher = {IOP Publishing},
volume = {15},
number = {11},
pages = {113022},
author = {Rossi, M and Huber, M and Bruß, D and Macchiavello, C},
title = {Quantum hypergraph states},
journal = {New Journal of Physics}
}

@article{bravyi2019simulationquantumcircuitslowrank,
  doi = {10.22331/q-2019-09-02-181},
  url = {https://doi.org/10.22331/q-2019-09-02-181},
  title = {Simulation of quantum circuits by low-rank stabilizer decompositions},
  author = {Bravyi, Sergey and Browne, Dan and Calpin, Padraic and Campbell, Earl and Gosset, David and Howard, Mark},
  journal = {{Quantum}},
  issn = {2521-327X},
  publisher = {{Verein zur F{\"{o}}rderung des Open Access Publizierens in den Quantenwissenschaften}},
  volume = {3},
  pages = {181},
  month = sep,
  year = {2019}
}

@article{aaronson2008improvedsimulationstabilizercircuits,
  title = {Improved simulation of stabilizer circuits},
  author = {Aaronson, Scott and Gottesman, Daniel},
  journal = {Phys. Rev. A},
  volume = {70},
  issue = {5},
  pages = {052328},
  numpages = {14},
  year = {2004},
  month = {Nov},
  publisher = {American Physical Society},
  doi = {10.1103/PhysRevA.70.052328},
  url = {https://link.aps.org/doi/10.1103/PhysRevA.70.052328}
}

@Article{Williamson2026,
  author    = {Williamson, Dominic J. and Yoder, Theodore J.},
  journal   = {Nature Physics},
  title     = {Low-overhead fault-tolerant quantum computation by gauging logical operators},
  year      = {2026},
  issn      = {1745-2481},
  month     = Apr,
  number    = {4},
  pages     = {598--603},
  volume    = {22},
  doi       = {10.1038/s41567-026-03220-8},
  publisher = {Springer Science and Business Media LLC},
}

@misc{guemard2025liftingcsscodehandlebody,
      title={Lifting a CSS code via its handlebody realization}, 
      author={Virgile Guemard},
      year={2025},
      eprint={2505.14327},
      archivePrefix={arXiv},
      primaryClass={quant-ph},
      url={https://arxiv.org/abs/2505.14327}, 
}

@misc{wills2026forcedgappostselectionquantum,
      title={Forced Gap Post-Selection for Quantum LDPC Codes and their Operations}, 
      author={Adam Wills and Theodore J. Yoder and Isaac Chuang},
      year={2026},
      eprint={2605.20346},
      archivePrefix={arXiv},
      primaryClass={quant-ph},
      url={https://arxiv.org/abs/2605.20346}, 
}

@inproceedings{Zhou_2025, series={SIGARCH ’25},
   title={Resource Analysis of Low-Overhead Transversal Architectures for Reconfigurable Atom Arrays},
   url={http://dx.doi.org/10.1145/3695053.3731039},
   DOI={10.1145/3695053.3731039},
   booktitle={Proceedings of the 52nd Annual International Symposium on Computer Architecture},
   publisher={ACM},
   author={Zhou, Hengyun and Duckering, Casey and Zhao, Chen and Bluvstein, Dolev and Cain, Madelyn and Kubica, Aleksander and Wang, Sheng-Tao and Lukin, Mikhail D.},
   year={2025},
   month=jun, pages={1432–1448},
   collection={SIGARCH ’25} }

@misc{leverrier2026approximatingoptimaldecodingquantum,
      title={Approximating optimal decoding of quantum LDPC codes with narrow frontiers}, 
      author={Anthony Leverrier and Rüdiger Urbanke},
      year={2026},
      eprint={2606.20513},
      archivePrefix={arXiv},
      primaryClass={quant-ph},
      url={https://arxiv.org/abs/2606.20513}, 
}

@article{PhysRevLett.111.090505,
  title = {Universal Fault-Tolerant Quantum Computation with Only Transversal Gates and Error Correction},
  author = {Paetznick, Adam and Reichardt, Ben W.},
  journal = {Phys. Rev. Lett.},
  volume = {111},
  issue = {9},
  pages = {090505},
  numpages = {5},
  year = {2013},
  month = {Aug},
  publisher = {American Physical Society},
  doi = {10.1103/PhysRevLett.111.090505},
  url = {https://link.aps.org/doi/10.1103/PhysRevLett.111.090505}
}

@article{Haah2018codesprotocols,
  doi = {10.22331/q-2018-06-07-71},
  url = {https://doi.org/10.22331/q-2018-06-07-71},
  title = {Codes and {P}rotocols for {D}istilling {$T$}, controlled-{$S$}, and {T}offoli {G}ates},
  author = {Haah, Jeongwan and Hastings, Matthew B.},
  journal = {{Quantum}},
  issn = {2521-327X},
  publisher = {{Verein zur F{\"{o}}rderung des Open Access Publizierens in den Quantenwissenschaften}},
  volume = {2},
  pages = {71},
  month = jun,
  year = {2018}
}

@article{PRXQuantum.4.020332,
  title = {Improved Single-Shot Decoding of Higher-Dimensional Hypergraph-Product Codes},
  author = {Higgott, Oscar and Breuckmann, Nikolas P.},
  journal = {PRX Quantum},
  volume = {4},
  issue = {2},
  pages = {020332},
  numpages = {17},
  year = {2023},
  month = {May},
  publisher = {American Physical Society},
  doi = {10.1103/PRXQuantum.4.020332},
  url = {https://link.aps.org/doi/10.1103/PRXQuantum.4.020332}
}

@article{PRXQuantum.5.010302,
  title = {Fault-Tolerant Postselection for Low-Overhead Magic State Preparation},
  author = {Bomb{\'{\i}}n, H{\'e}ctor and Pant, Mihir and Roberts, Sam and Seetharam, Karthik I.},
  journal = {PRX Quantum},
  volume = {5},
  issue = {1},
  pages = {010302},
  numpages = {19},
  year = {2024},
  month = {Jan},
  publisher = {American Physical Society},
  doi = {10.1103/PRXQuantum.5.010302},
  url = {https://link.aps.org/doi/10.1103/PRXQuantum.5.010302}
}

@misc{meister2024efficientsoftoutputdecoderssurface,
      title={Efficient soft-output decoders for the surface code},
      author={Nadine Meister and Christopher A. Pattison and John Preskill},
      year={2024},
      eprint={2405.07433},
      archivePrefix={arXiv},
      primaryClass={quant-ph},
      url={https://arxiv.org/abs/2405.07433},
}

@article{PhysRevA.89.022326,
  title = {Efficient Markov chain Monte Carlo algorithm for the surface code},
  author = {Hutter, Adrian and Wootton, James R. and Loss, Daniel},
  journal = {Phys. Rev. A},
  volume = {89},
  issue = {2},
  pages = {022326},
  numpages = {10},
  year = {2014},
  month = {Feb},
  publisher = {American Physical Society},
  doi = {10.1103/PhysRevA.89.022326},
  url = {https://link.aps.org/doi/10.1103/PhysRevA.89.022326}
}

@article{vaknin2026efficientmagicstatecultivation,
  title = {High Rate Magic State Cultivation on the Surface Code},
  author = {Vaknin, Yotam and Jacoby, Shoham and Grimsmo, Arne and Retzker, Alex},
  journal = {PRX Quantum},
  volume = {7},
  issue = {1},
  pages = {010353},
  numpages = {20},
  year = {2026},
  month = {Mar},
  publisher = {American Physical Society},
  doi = {10.1103/p8tw-6kq9},
  url = {https://link.aps.org/doi/10.1103/p8tw-6kq9}
}

@article{PRXQuantum.2.020340,
  title = {Single-Shot Error Correction of Three-Dimensional Homological Product Codes},
  author = {Quintavalle, Armanda O. and Vasmer, Michael and Roffe, Joschka and Campbell, Earl T.},
  journal = {PRX Quantum},
  volume = {2},
  issue = {2},
  pages = {020340},
  numpages = {32},
  year = {2021},
  month = {Jun},
  publisher = {American Physical Society},
  doi = {10.1103/PRXQuantum.2.020340},
  url = {https://link.aps.org/doi/10.1103/PRXQuantum.2.020340}
}

@article{PhysRevX.5.031043,
  title = {Single-Shot Fault-Tolerant Quantum Error Correction},
  author = {Bomb{\'{\i}}n, H{\'e}ctor},
  journal = {Phys. Rev. X},
  volume = {5},
  issue = {3},
  pages = {031043},
  numpages = {26},
  year = {2015},
  month = {Sep},
  publisher = {American Physical Society},
  doi = {10.1103/PhysRevX.5.031043},
  url = {https://link.aps.org/doi/10.1103/PhysRevX.5.031043}
}

@article{Gidney2019efficientmagicstate,
  doi = {10.22331/q-2019-04-30-135},
  url = {https://doi.org/10.22331/q-2019-04-30-135},
  title = {Efficient magic state factories with a catalyzed {$|CCZ\rangle$} to {$2|T\rangle$} transformation},
  author = {Gidney, Craig and Fowler, Austin G.},
  journal = {{Quantum}},
  issn = {2521-327X},
  publisher = {{Verein zur F{\"{o}}rderung des Open Access Publizierens in den Quantenwissenschaften}},
  volume = {3},
  pages = {135},
  month = apr,
  year = {2019}
}

@article{PhysRevA.95.022316,
  title = {Unified framework for magic state distillation and multiqubit gate synthesis with reduced resource cost},
  author = {Campbell, Earl T. and Howard, Mark},
  journal = {Phys. Rev. A},
  volume = {95},
  issue = {2},
  pages = {022316},
  numpages = {22},
  year = {2017},
  month = {Feb},
  publisher = {American Physical Society},
  doi = {10.1103/PhysRevA.95.022316},
  url = {https://link.aps.org/doi/10.1103/PhysRevA.95.022316}
}

@article{PhysRevA.95.032338,
  title = {Quantum computation with realistic magic-state factories},
  author = {O'Gorman, Joe and Campbell, Earl T.},
  journal = {Phys. Rev. A},
  volume = {95},
  issue = {3},
  pages = {032338},
  numpages = {19},
  year = {2017},
  month = {Mar},
  publisher = {American Physical Society},
  doi = {10.1103/PhysRevA.95.032338},
  url = {https://link.aps.org/doi/10.1103/PhysRevA.95.032338}
}

@article{Gidney2021howtofactorbit,
  doi = {10.22331/q-2021-04-15-433},
  url = {https://doi.org/10.22331/q-2021-04-15-433},
  title = {How to factor 2048 bit {RSA} integers in 8 hours using 20 million noisy qubits},
  author = {Gidney, Craig and Eker{\aa{}}, Martin},
  journal = {{Quantum}},
  issn = {2521-327X},
  publisher = {{Verein zur F{\"{o}}rderung des Open Access Publizierens in den Quantenwissenschaften}},
  volume = {5},
  pages = {433},
  month = apr,
  year = {2021}
}

@article{PhysRevA.87.022328,
  title = {Low-overhead constructions for the fault-tolerant Toffoli gate},
  author = {Jones, Cody},
  journal = {Phys. Rev. A},
  volume = {87},
  issue = {2},
  pages = {022328},
  numpages = {4},
  year = {2013},
  month = {Feb},
  publisher = {American Physical Society},
  doi = {10.1103/PhysRevA.87.022328},
  url = {https://link.aps.org/doi/10.1103/PhysRevA.87.022328}
}

@article{PhysRevLett.118.090501,
  title = {Application of a Resource Theory for Magic States to Fault-Tolerant Quantum Computing},
  author = {Howard, Mark and Campbell, Earl},
  journal = {Phys. Rev. Lett.},
  volume = {118},
  issue = {9},
  pages = {090501},
  numpages = {6},
  year = {2017},
  month = {Mar},
  publisher = {American Physical Society},
  doi = {10.1103/PhysRevLett.118.090501},
  url = {https://link.aps.org/doi/10.1103/PhysRevLett.118.090501}
}

@misc{bombin20182dquantumcomputation3d,
      title={2D quantum computation with 3D topological codes}, 
      author={Hector Bombin},
      year={2018},
      eprint={1810.09571},
      archivePrefix={arXiv},
      primaryClass={quant-ph},
      url={https://arxiv.org/abs/1810.09571} 
}

@misc{kim2026xtim,
  author       = {Kim, Isaac H.},
  title        = {{xtim}},
  year         = {2026},
  month        = nov,
  howpublished = {\url{https://github.com/ikim-quantum/xtim}},
  note         = {Version 4.1.2}
}

@misc{hirasaki2026cczfactorycode,
  howpublished = {\url{https://github.com/y-hirasaki/ExactSimulation_CCZfactory}}
}

@Article{Ruiz2026,
  author    = {Ruiz, Diego and Guillaud, Jérémie and Vuillot, Christophe and Mirrahimi, Mazyar},
  journal   = {npj Quantum Information},
  title     = {Unfolded distillation: very low-cost magic state preparation for biased-noise qubits},
  year      = {2026},
  issn      = {2056-6387},
  month     = Feb,
  number    = {1},
  volume    = {12},
  doi       = {10.1038/s41534-026-01197-z},
  publisher = {Springer Science and Business Media LLC},
}

@Article{Brown2020,
  author    = {Brown, Benjamin J.},
  journal   = {Science Advances},
  title     = {A fault-tolerant non-Clifford gate for the surface code in two dimensions},
  year      = {2020},
  issn      = {2375-2548},
  month     = May,
  number    = {21},
  volume    = {6},
  doi       = {10.1126/sciadv.aay4929},
  publisher = {American Association for the Advancement of Science (AAAS)},
}

@misc{christos2026nonabelianquantumlowdensityparity,
      title={Non-Abelian Quantum Low-Density Parity Check Codes and Non-Clifford Operations from Gauging Logical Gates via Measurements}, 
      author={Maine Christos and Chiu Fan Bowen Lo and Vedika Khemani and Rahul Sahay},
      year={2026},
      eprint={2602.12228},
      archivePrefix={arXiv},
      primaryClass={quant-ph},
      url={https://arxiv.org/abs/2602.12228}, 
}

@misc{zhu2026nonabelianqldpctqftformalism,
      title={Non-Abelian qLDPC: TQFT Formalism, Addressable Gauging Measurement and Application to Magic State Fountain on 2D Product Codes}, 
      author={Guanyu Zhu and Ryohei Kobayashi and Po-Shen Hsin},
      year={2026},
      eprint={2601.06736},
      archivePrefix={arXiv},
      primaryClass={quant-ph},
      url={https://arxiv.org/abs/2601.06736}, 
}

@misc{chang2026magicquantumcodesurgery,
      title={Magic Quantum Code Surgery}, 
      author={Kathleen Chang and Anasuya Lyons and Yuanjie Ren and Harald Putterman and Nathanan Tantivasadakarn and Victor V. Albert and Benjamin J. Brown and Dominic J. Williamson},
      year={2026},
      eprint={2610.03920},
      archivePrefix={arXiv},
      primaryClass={quant-ph},
      url={https://arxiv.org/abs/2610.03920}, 
}

@misc{williamson2026fastmagicstatepreparation,
      title={Fast magic state preparation by gauging higher-form transversal gates in parallel}, 
      author={Dominic J. Williamson},
      year={2026},
      eprint={2601.22939},
      archivePrefix={arXiv},
      primaryClass={quant-ph},
      url={https://arxiv.org/abs/2601.22939}, 
}

@misc{li2026nonabeliansheafquantumldpc,
      title={Non-Abelian sheaf quantum LDPC codes: good and magical}, 
      author={Zimu Li and Fuchuan Wei and Zhengyi Han and Zi-Wen Liu},
      year={2026},
      eprint={2609.30159},
      archivePrefix={arXiv},
      primaryClass={quant-ph},
      url={https://arxiv.org/abs/2609.30159}, 
}

@Article{breuckmann2024cups,
author="Breuckmann, Nikolas P.
and Davydova, Margarita
and Eberhardt, Jens N.
and Tantivasadakarn, Nathanan",
title="Cups and Gates {I}: Cohomology Invariants and Logical Quantum Operations",
journal="Communications in Mathematical Physics",
year="2026",
month="Apr",
day="04",
volume="407",
number="5",
pages="86",
issn="1432-0916",
doi="10.1007/s00220-026-05570-z",
url="https://doi.org/10.1007/s00220-026-05570-z"
}

@misc{breuckmann2026cupsgatesiioperads,
      title={Cups and Gates {II}: Operads, twists, and addressable intra-code gates},
      author={Nikolas P. Breuckmann and Margarita Davydova and Jens Niklas Eberhardt and Lucas Ferdinand Hallwirth and Nathanan Tantivasadakarn},
      year={2026},
      eprint={2610.05214},
      archivePrefix={arXiv},
      primaryClass={quant-ph},
      url={https://arxiv.org/abs/2610.05214},
}

@misc{hsin2025classifyinglogicalgatesquantum,
      title={Classifying Logical Gates in Quantum Codes via Cohomology Operations and Symmetry},
      author={Po-Shen Hsin and Ryohei Kobayashi and Guanyu Zhu},
      year={2025},
      eprint={2411.15848},
      archivePrefix={arXiv},
      primaryClass={quant-ph},
      url={https://arxiv.org/abs/2411.15848},
}

@article{wcxs-w69t,
  title = {Non-{Clifford} and Parallelizable Fault-Tolerant Logical Gates on Constant and Almost-Constant Rate Homological Quantum Low-Density Parity-Check Codes via Higher Symmetries},
  author = {Zhu, Guanyu and Sikander, Shehryar and Portnoy, Elia and Cross, Andrew W. and Brown, Benjamin J.},
  journal = {PRX Quantum},
  volume = {6},
  issue = {4},
  pages = {040361},
  numpages = {43},
  year = {2025},
  month = {Dec},
  publisher = {American Physical Society},
  doi = {10.1103/wcxs-w69t},
  url = {https://link.aps.org/doi/10.1103/wcxs-w69t}
}

@misc{zhu2025topologicaltheoryqldpcnonclifford,
      title={A topological theory for {qLDPC}: non-{Clifford} gates and magic state fountain on homological product codes with constant rate and beyond the {$N^{1/3}$} distance barrier},
      author={Guanyu Zhu},
      year={2025},
      eprint={2501.19375},
      archivePrefix={arXiv},
      primaryClass={quant-ph},
      url={https://arxiv.org/abs/2501.19375},
}

@misc{zhu2025transversalnoncliffordgatesqldpc,
      title={Transversal non-{Clifford} gates on {qLDPC} codes breaking the {$\sqrt{N}$} distance barrier and quantum-inspired geometry with {$\mathbb{Z}_2$} systolic freedom},
      author={Guanyu Zhu},
      year={2025},
      eprint={2507.15056},
      archivePrefix={arXiv},
      primaryClass={quant-ph},
      url={https://arxiv.org/abs/2507.15056},
}

@misc{tiew2026copycupgatestensorproducts,
      title={Copy-cup Gates in Tensor Products of Group Algebra Codes},
      author={Ryan Tiew and Nikolas P. Breuckmann},
      year={2026},
      eprint={2602.23307},
      archivePrefix={arXiv},
      primaryClass={quant-ph},
      url={https://arxiv.org/abs/2602.23307},
}

@misc{lin2024transversalnoncliffordgatesquantum,
      title={Transversal non-{Clifford} gates for quantum {LDPC} codes on sheaves},
      author={Ting-Chun Lin},
      year={2024},
      eprint={2410.14631},
      archivePrefix={arXiv},
      primaryClass={quant-ph},
      url={https://arxiv.org/abs/2410.14631},
}

@misc{li2026theorycohomologicalinvariantsquantum,
      title={Theory of (Co)homological Invariants on Quantum {LDPC} Codes},
      author={Zimu Li and Yuguo Shao and Fuchuan Wei and Yiming Li and Zi-Wen Liu},
      year={2026},
      eprint={2603.25831},
      archivePrefix={arXiv},
      primaryClass={quant-ph},
      url={https://arxiv.org/abs/2603.25831},
}

@misc{li2026transversalnoncliffordgatesalmostgood,
      title={Transversal non-{Clifford} gates on almost-good quantum {LDPC} and quantum locally testable codes},
      author={Yiming Li and Zimu Li and Zi-Wen Liu},
      year={2026},
      eprint={2604.01874},
      archivePrefix={arXiv},
      primaryClass={quant-ph},
      url={https://arxiv.org/abs/2604.01874},
}

@misc{li2026parallelizableaddressabletransversalnonclifford,
      title={Parallelizable and addressable transversal non-{Clifford} gates on good quantum {LDPC} codes},
      author={Yiming Li and Zimu Li and Zhengyi Han and Zi-Wen Liu},
      year={2026},
      eprint={2609.26691},
      archivePrefix={arXiv},
      primaryClass={quant-ph},
      url={https://arxiv.org/abs/2609.26691},
}

@misc{golowich2026improvedtransversalnoncliffordgates,
      title={Improved Transversal Non-{Clifford} Gates from Cup Products},
      author={Louis Golowich and Itzhak Tamo and Guanyu Zhu},
      year={2026},
      eprint={2609.27801},
      archivePrefix={arXiv},
      primaryClass={quant-ph},
      url={https://arxiv.org/abs/2609.27801},
}

@misc{jacob2026singleshotdecodingfaulttolerantgates,
      title={Single-Shot Decoding and Fault-tolerant Gates with Trivariate Tricycle Codes},
      author={Abraham Jacob and Campbell McLauchlan and Dan E. Browne},
      year={2026},
      eprint={2508.08191},
      archivePrefix={arXiv},
      primaryClass={quant-ph},
      url={https://arxiv.org/abs/2508.08191},
}

@article{ghhp-cytl,
  title = {Magic Tricycles: Efficient Magic-State Generation with Finite Block-Length Quantum {LDPC} Codes},
  author = {Menon, Varun and Bonilla Ataides, J. Pablo and Mehta, Rohan and Gu, Andi and Tan, Daniel Bochen and Lukin, Mikhail D.},
  journal = {Phys. Rev. X},
  volume = {16},
  issue = {2},
  pages = {021014},
  numpages = {37},
  year = {2026},
  month = {Apr},
  publisher = {American Physical Society},
  doi = {10.1103/ghhp-cytl},
  url = {https://link.aps.org/doi/10.1103/ghhp-cytl}
}

@misc{li2026transversaldimensionjumpproduct,
      title={Transversal dimension jump for product {qLDPC} codes},
      author={Christine Li and John Preskill and Qian Xu},
      year={2026},
      eprint={2510.07269},
      archivePrefix={arXiv},
      primaryClass={quant-ph},
      url={https://arxiv.org/abs/2510.07269},
}

@misc{tan2025singleshotuniversalityquantumldpc,
      title={Single-Shot Universality in Quantum {LDPC} Codes via Code-Switching},
      author={Shi Jie Samuel Tan and Yifan Hong and Ting-Chun Lin and Michael J. Gullans and Min-Hsiu Hsieh},
      year={2025},
      eprint={2510.08552},
      archivePrefix={arXiv},
      primaryClass={quant-ph},
      url={https://arxiv.org/abs/2510.08552},
}

@inproceedings{golowich2024quantumldpccodestransversal,
  author = {Louis Golowich and Ting-Chun Lin},
  title = {Quantum {LDPC} Codes with Transversal Non-{Clifford} Gates via Products of Algebraic Codes},
  booktitle = {Proceedings of the 57th Annual ACM Symposium on Theory of Computing},
  publisher = {Association for Computing Machinery},
  year = {2025},
  pages = {689--696},
  doi = {10.1145/3717823.3718139},
  eprint = {2410.14662},
  archivePrefix = {arXiv},
  primaryClass = {quant-ph}
}

\end{document}